\documentclass[floatfix, preprint, showpacs, showkeys, preprintnumbers, nofootinbib, superscriptaddress]{revtex4-1}
\usepackage[utf8]{inputenc}
\usepackage[sort&compress]{natbib}
\usepackage{ulem}
\usepackage{bm}
\usepackage{times}
\usepackage{amssymb,amsbsy,amsmath,amsfonts}
\usepackage{graphicx}
\usepackage{epstopdf}
\usepackage{float}
\usepackage{color}
\usepackage{morefloats}
\usepackage{rotating}
\usepackage{srcltx}
\usepackage{slashed}
\usepackage{subfigure}
\usepackage{multirow}
\usepackage{verbatim}
\usepackage{hyperref}
\usepackage{overpic}
\usepackage{diagbox}
\usepackage{tabularx}

\begin{document}

\title{Search for hidden-charm pentaquark states in  three-body final states }

\author{Jia-Ming Xie}
\affiliation{School of Physics, Beihang University, Beijing 100191, China}

\author{Xi-Zhe Ling}
\affiliation{Institute of High Energy Physics,  Chinese Academy of Sciences, Beijing 100049, China  }

\author{Ming-Zhu Liu}\email{zhengmz11@buaa.edu.cn}
\affiliation{School of Space and Environment, Beihang University, Beijing 100191, China}

\author{Li-Sheng Geng}\email{lisheng.geng@buaa.edu.cn}
\affiliation{School of Physics, Beihang University, Beijing 100191, China}
\affiliation{
Beijing Key Laboratory of Advanced Nuclear Materials and Physics,
Beihang University, Beijing 100191, China}
\affiliation{School of Physics and Microelectronics, Zhengzhou University, Zhengzhou, Henan 450001, China}

\date{\today}
\begin{abstract}

    The three pentaquark states,  $P_c(4312)$, $P_c(4440)$ and $P_c(4457)$,  discovered by the LHCb Collaboration in 2019,  are often interpreted as $\bar{D}^{(\ast)}\Sigma_c$ molecules. Together with their four $\bar{D}^{(*)}\Sigma_{c}^{\ast}$ partners dictated by heavy quark spin symmetry they represent a complete multiplet of hadronic molecules of $\bar{D}^{(\ast)}\Sigma_{c}^{(\ast)}$. The pentaquark states were observed in the $J/\psi p$ invariant mass distributions of the $\Lambda_b\rightarrow J/\psi p K$ decay. It is widely recognized that to understand their nature, other discovery channels play an important role. In this work, we investigate two  three-body decay modes of the $\bar{D}^{(\ast)}\Sigma_{c}^{(\ast)}$ molecules.  The tree-level modes proceed via  off-shell $\Sigma_{c}^{(\ast)}$ baryons, $\bar{D}^{(\ast)}\Sigma_{c}^{(\ast)} \to \bar{D}^{(\ast)}\left(\Sigma_{c}^{(\ast)}\to \Lambda_{c}\pi\right)\to\bar{D}^{(\ast)}\Lambda_{c}\pi$, while the triangle-loop modes  proceed through $\bar{D}^{\ast}\Sigma_{c}^{(\ast)}\to J/\psi N\pi$, $\eta_{c}N\pi$ via $\bar{D}\Sigma_{c}^{(\ast)}$  rescattering to $J/\psi N$ and $\eta_{c}N$. 
Our results indicate  that the decay widths of the  $P_{c}(4457)$ and $\bar{D}^{(\ast)}\Sigma_{c}^{\ast}$ states  into $\bar{D}^{(\ast)}\Lambda_{c}\pi$ are several MeV, as a result can be observed in the upcoming Run 3 and Run 4 of LHC. The partial decay widths into $\bar{D}^{(\ast)}\Lambda_{c}\pi$ of the $P_{c}(4312)$ and $P_{c}(4440)$ states range from tens to hundreds of keV. In addition, the partial decay widths of $\bar{D}^{\ast}\Sigma_{c}$ molecules into $J/\psi N \pi$ and $\eta_c N \pi$ are several keV and tens of keV, respectively, and the partial decay widths of $\bar{D}^{\ast}\Sigma_{c}^{\ast}$ molecules  into $J/\psi N \pi$ vary from several keV to tens of keV. In particular, we show that the spin-5/2 $\bar{D}^{\ast}\Sigma_{c}^{\ast}$ state can be searched for in the $J/\psi N \pi$ and $\bar{D}^{\ast}\Lambda_{c}\pi$ invariant mass distributions, while the latter one is more favorable.
These three-body  decay modes of the pentaquark states  are of great value to further observations of the pentaquark states and to  a better understanding of their  nature.

\end{abstract}


\maketitle

\section{Introduction}

The existence of hidden-charm pentaquark states was  predicted about ten years ago~\cite{Wu:2010jy,Wu:2010vk,Wang:2011rga,Yang:2011wz,Yuan:2012wz,Wu:2012md,Garcia-Recio:2013gaa,Xiao:2013yca,Uchino:2015uha,Karliner:2015ina}.  In 2015, the LHCb Collaboration  observed  two pentaquark states, named as $P_{c}(4380)$ and $P_{c}(4450)$, in the $J/\psi p$ invariant mass distributions  of  the $\Lambda_{b}\to J/\psi p K$ decay~\cite{Aaij:2015tga}.  In 2019, the LHCb Collaboration updated their analysis with ten times more data, showing that the original $P_{c}(4450)$ state splits into two states, $P_{c}(4440)$ and $P_{c}(4457)$, and in addition a new state $P_{c}(4312)$ is discovered~\cite{Aaij:2019vzc}.  The masses and decay widths of the three  states are 
\begin{eqnarray}
 \nonumber
M_{P_{c}(4312)}&=&4311.9\pm 0.7^{+6.8}_{-0.6} ~\mbox{MeV}    \quad \quad   \Gamma_{P_{c}(4312)}=9.8 \pm 2.7^{+3.7}_{-4.5} ~\mbox{MeV},   \\
M_{P_{c}(4440)}&=&4440.3\pm 1.3^{+4.1}_{-4.7} ~\mbox{MeV}    \quad \quad   \Gamma_{P_{c}(4440)}=20.6 \pm 4.9^{+8.7}_{-10.1}~\mbox{MeV},
\\ \nonumber
M_{P_{c}(4457)}&=&4457.3\pm 0.6^{+4.1}_{-1.7} ~\mbox{MeV}    \quad \quad   \Gamma_{P_{c}(4457)}=6.4 \pm 2.0^{+5.7}_{-1.9} ~\mbox{MeV}.
\end{eqnarray}

Two more pentaquark states were reported in the following years, though only with a significance of about 3$\sigma$. A hidden-charm pentaquark with strangeness, $P_{cs}(4459)$, is  visible  in the $J/\psi \Lambda$ invariant mass spectrum~\cite{LHCb:2020jpq}, and a hidden-charm  pentaquark $P_{c}(4337)$ was found in both the $J/\psi p$ and $J/\psi \bar{p}$ invariant mass spectrum~\cite{LHCb:2021chn}. The former has been predicted by many studies as the $SU(3)$-flavor partner of the pentaquark states, $P_{c}(4312)$, $P_{c}(4380)$,  $P_{c}(4440)$, and $P_{c}(4457)$~\cite{Chen:2015sxa,Chen:2016ryt,Shen:2019evi,Xiao:2019gjd,Wang:2019nvm}, while the latter is more difficult to understand. It could be a $\chi_{c0}(1S) p$ bound state~\cite{Yan:2021nio},  a compact multiquark state~\cite{Deng:2022vkv},  a cusp effect~\cite{Nakamura:2021dix}, or a reflection effect~\cite{Wang:2021crr}.   Therefore,  we will leave a study of $P_{cs}(4459)$ and $P_{c}(4337)$ to a future work.          

Trying to understand the nature of the pentaquark states has led to intensive theoretical  studies.  In Refs.~\cite{Liu:2019tjn,Liu:2019zvb} we have employed both an effective field theory (EFT) and the one-boson-exchange (OBE) model to describe the three pentaquark states as $\bar{D}^{(\ast)}\Sigma_{c}$ molecules by reproducing their masses, which is later  confirmed by many other theoretical studies~\cite{Chen:2019asm,He:2019ify,Chen:2019bip,Xiao:2019aya,Yamaguchi:2019seo,Valderrama:2019chc,Du:2019pij,He:2019rva,Wang:2019spc}. In addition to the $\bar{D}^{(\ast)}\Sigma_{c}$ channel,  the role of  the $\bar{D}\Lambda_{c1}$ channel has been studied in relation with the $P_c(4457)$ state~\cite{Burns:2019iih,Peng:2020gwk,Yalikun:2021bfm}.  In Refs.~\cite{Xiao:2019mvs,Lin:2019qiv} with the effective Lagrangian approach the authors   reproduced the decay widths of the pentaquark states in the hadronic molecular picture.  With the same approach, Wu \begin{itshape}et al.\end{itshape} calculated the ratios of  the production rates of the pentaquark states in the $\Lambda_{b}$ decays, and obtained results  in
agreement with the LHCb measurements~\cite{Wu:2019rog}. Although the molecular interpretation is the most popular, other explanations are also available, e.g., hadro-charmonia~\cite{Eides:2019tgv}, compact pentaquark states~\cite{Ali:2019npk,Mutuk:2019snd,Wang:2019got,Cheng:2019obk,Weng:2019ynv,Zhu:2019iwm,Pimikov:2019dyr,Ruangyoo:2021aoi,Deng:2022vkv}, 
virtual states~\cite{Fernandez-Ramirez:2019koa} or double triangle singularities~\cite{Nakamura:2021qvy}. 

To  investigate the molecular nature of the pentaquark states, many other methods have been proposed.  Historically, the existence of the $\Omega$ baryon indeed verified the quark model, where the $SU(3)$-flavor symmetry plays  an important role.  Following the same  principle, 
  among others, we  proposed that, if future experiments could discover the four heavy quark spin symmetry (HQSS) partners of the $\bar{D}^{(\ast)}\Sigma_{c}^{\ast}$  molecules, it would support their molecular nature~\cite{Liu:2019tjn}. Apart from HQSS, their $SU(3)$-flavor symmetry partners~\cite{Xiao:2019gjd,Wang:2019nvm} and heavy quark flavor symmetry partners~\cite{Wang:2019ato,Azizi:2020ogm} have been predicted, the existence of which can also  support the molecular nature of the pentaquark states. In our previous work~\cite{Pan:2019skd}, we showed that one can study instead the $\Xi_{cc}^{(\ast)}\Sigma_{c}^{(\ast)}$ system, because it can be related to the $\bar{D}^{(\ast)}\Sigma_{c}^{(\ast)}$ system via  heavy antiquark diquark symmetry (HADS). In addition,  lattice QCD simulations can provide valuable information   to understand the pentaquark states, but a complete simulation of the pentaquark system with all the relevant coupled channels is complicated~\cite{Skerbis:2018lew,Sugiura:2019pye}.  Recently, we predicted the existence of a three-body
hadronic molecule $\Sigma_{c}\bar{D}\bar{K}$~\cite{Wu:2021gyn}, which can be viewed as an excited state of the $P_{cs}(4459)$  state.  
If the $\Sigma_{c}\bar{D}\bar{K}$  molecule is  discovered in the future, it will provide a non-trivial check on the molecular nature of the  pentaquark states.   

The decay and production mechanisms  of the pentaquark states have also attracted considerable attention, which can offer valuable means  to reveal their nature. 
In Ref.~\cite{Sakai:2019qph}, assuming the pentaquark states as hadronic molecules,  Sakai \begin{itshape}et al.\end{itshape} found that the decay width of   $P_{c}(4312)\to \eta_{c} p$ is about three times larger than that of  $ P_{c}(4312)\to J/\psi p $, in agreement with the results of the quark interchange model~\cite{Wang:2019spc}. However, if $P_{c}(4312)$ is regarded as a compact pentaquark state, it will dominantly decay into $\bar{D}^{\ast}\Lambda_{c}$ rather than $\eta_{c}p$~\cite{Weng:2019ynv}. In Ref.\cite{Guo:2019fdo}, Guo \begin{itshape}et al.\end{itshape} estimated that the branching ratio  Br$(P_{c}(4457)\to J/\psi \Delta)$/Br$(P_{c}(4457)\to J/\psi p)$ ranges from a few percent to about $30\%$ in the molecular picture, which shows a larger isospin breaking in comparison with the decays of ordinary hadrons.    As a result, experimental measurements of such branching ratios are of great value  to reveal the internal structure  of the pentaquark states. 
In Ref.~\cite{Chen:2022dmr},  Chen \begin{itshape}et al.\end{itshape} estimated  the production rates of $P_{c}(4312)$, $P_{c}(4440)$ and $P_{c}(4457)$ in $pp$ collisions using the constrained phasespace coalescence model and the parton and hadron cascade (PACIAE) model. Their results show that the production rates of  the pentaquark states, the nucleuslike states, and the hadronic molecular states are of the same order. 
In Ref.~\cite{Yang:2021jof}, Yang \begin{itshape}et al.\end{itshape} estimated the production rates of $P_{c}$'s in lepton-proton processes, where the $P_{c}$'s are treated as hadronic molecules.   

To further explore the decay mechanism of the pentaquark states in the molecular picture,  in this work  we study the three-body decays of the seven $\bar{D}^{(\ast)}\Sigma_{c}^{(\ast)}$ pentaquark states. The decay mechanism of the $\bar{D}^{(\ast)}\Sigma_{c}^{(\ast)}$ molecules is similar to that of the $T_{cc}$ state recently discovered by the LHCb Collaboration~\cite{LHCb:2021vvq,LHCb:2021auc}, where  the decay of $T_{cc}$ in the molecular picture  proceeds via the off-shell $D^{\ast}$ meson~\cite{Meng:2021jnw,Ling:2021bir,Chen:2021vhg,Feijoo:2021ppq,Yan:2021wdl,Fleming:2021wmk,Albaladejo:2021vln,Du:2021zzh,Mikhasenko:2022rrl}.  As for  the $\bar{D}^{(\ast)}\Sigma_{c}^{(\ast)}$ molecules, we will consider two three-body decay modes. The tree-level modes proceed via  off-shell $\Sigma_{c}^{(\ast)}$ baryons, i.e. $\bar{D}^{(\ast)}\Sigma_{c}^{(\ast)} \to \bar{D}^{(\ast)}\left(\Sigma_{c}^{(\ast)}\to \Lambda_{c}\pi\right)\to\bar{D}^{(\ast)}\Lambda_{c}\pi$, while the triangle-loop modes  proceed through $\bar{D}^{\ast}\Sigma_{c}^{(\ast)}\to J/\psi N\pi$, $\eta_{c}N\pi$ via $\bar{D}\Sigma_{c}^{(\ast)}$  rescattering to $J/\psi N$ or $\eta_{c}N$. We use the effective Lagrangian approach to estimate the decay widths of these two modes in this work.    We hope that these decay modes can  offer further insights into the nature  of the pentaquark states.

This paper is organized as follows. In Sec.~\ref{theoretical}, we introduce the coupled-channel contact-range EFT of $\eta_{c}N-J/\psi N-\bar{D}^{(\ast)}\Sigma_{c}^{(\ast)}$ for the pentaquark system as well as present the decay amplitudes of the $\bar{D}^{(\ast)}\Sigma_{c}^{(\ast)}$ molecules  via tree-level  and triangle-loop diagrams using the effective Lagrangian approach. In Sec.~\ref{results}, we present the mass spectrum of $\bar{D}^{(\ast)}\Sigma_{c}^{(\ast)}$, their relevant couplings and the  decay widths of the $\bar{D}^{(\ast)}\Sigma_{c}^{(\ast)}$ molecules into $\bar{D}^{(\ast)}\Lambda_{c}\pi$ and $J/\psi(\eta_c) N \pi$. 
Finally, this paper ends with a short summary in Sec.~\ref{sum}.

\section{Theoretical framework}
\label{theoretical}

\begin{figure}[ttt]
\begin{center}
\begin{tabular}{cccc}
\begin{minipage}[t]{0.23\linewidth}
\begin{center}
\begin{overpic}[scale=0.09]{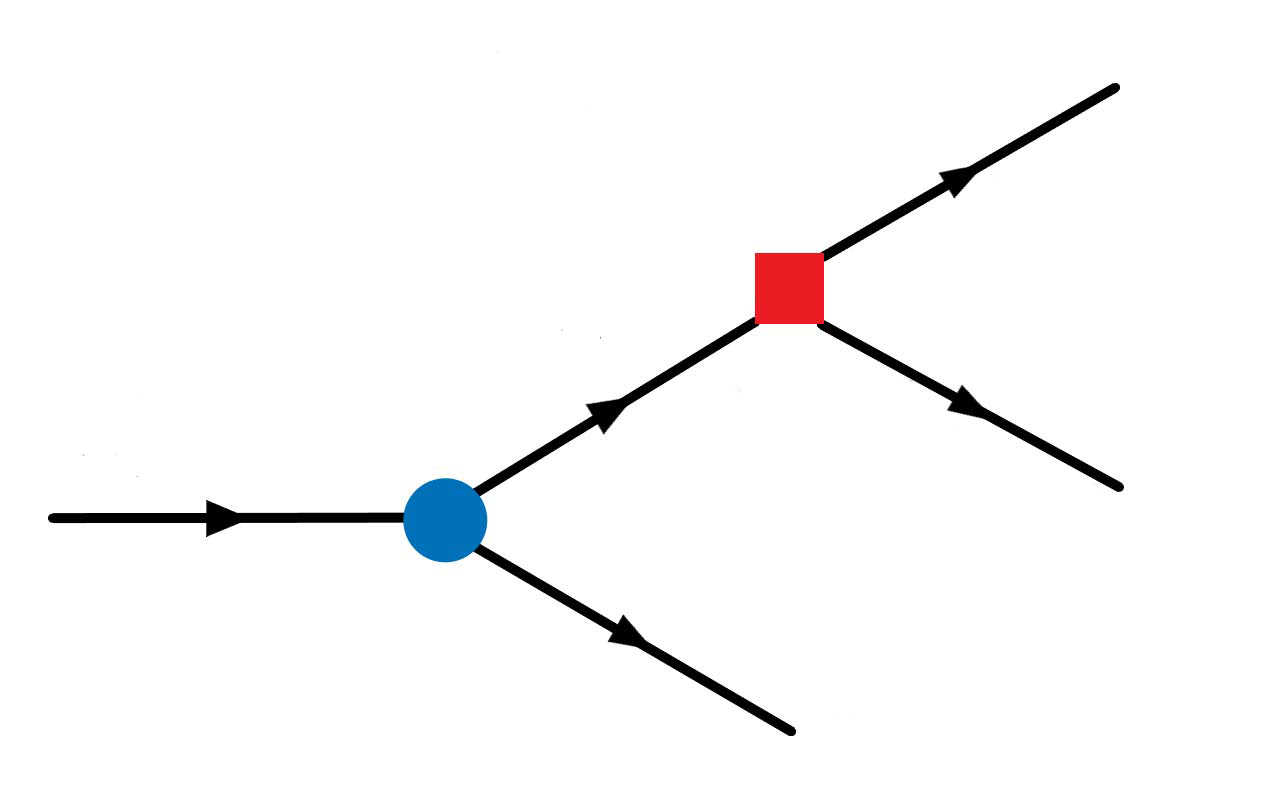}
		\put(12,26){$P_{c1}^+$}
		
		\put(40,3){$D^-$}
		
		\put(38,33){$\Sigma_{c}^{++}$}
		\put(68,52){$\Lambda_{c}^+$}
		\put(69,20){$\pi^+$}
		\put(45,-6){(a1)}
\end{overpic}
\end{center}
\end{minipage}
\hfill
\begin{minipage}[t]{0.23\linewidth}
\begin{center}
\begin{overpic}[scale=0.09]{tree.png}
		\put(12,26){$P_{c1}^+$}
		
		\put(38,3){$\bar{D}^0$}
		
		\put(39,33){$\Sigma_{c}^{+}$}
		\put(68,52){$\Lambda_{c}^+$}
		\put(69,20){$\pi^0$}
		\put(45,-6){(a2)}
\end{overpic}
\end{center}
\end{minipage}
\hfill
\begin{minipage}[t]{0.23\linewidth}
\begin{center}
\begin{overpic}[scale=0.09]{tree.png}
		\put(12,26){$P_{c2}^+$}
		
		\put(40,3){$D^-$}
		
		\put(35,33){$\Sigma_{c}^{\ast ++}$}
		\put(68,52){$\Lambda_{c}^+$}
		\put(69,20){$\pi^+$}
		\put(45,-6){(b1)}
\end{overpic}
\end{center}
\end{minipage}
\hfill
\begin{minipage}[t]{0.23\linewidth}
\begin{center}
\begin{overpic}[scale=0.09]{tree.png}
		\put(12,26){$P_{c2}^+$}
		
		\put(38,3){$\bar{D}^0$}
		
		\put(38,33){$\Sigma_{c}^{\ast +}$}
		\put(68,52){$\Lambda_{c}^+$}
		\put(69,20){$\pi^0$}
		\put(45,-6){(b2)}
\end{overpic}
\end{center}
\end{minipage}
\\
\begin{minipage}[t]{0.23\linewidth}
\begin{center}
\begin{overpic}[scale=0.09]{tree.png}
		\put(8,27){$P_{c3|c4}^+$}
		
		\put(36,3){$D^{\ast -}$}
		
		\put(38,33){$\Sigma_{c}^{++}$}
		\put(68,52){$\Lambda_{c}^+$}
		\put(69,20){$\pi^+$}
		\put(45,-6){(c1)}
\end{overpic}
\end{center}
\end{minipage}
\hfill
\begin{minipage}[t]{0.23\linewidth}
\begin{center}
\begin{overpic}[scale=0.09]{tree.png}
		\put(8,27){$P_{c3|c4}^+$}
		
		\put(35,3){$\bar{D}^{\ast 0}$}
		
		\put(39,33){$\Sigma_{c}^{+}$}
		\put(68,52){$\Lambda_{c}^+$}
		\put(69,20){$\pi^0$}
		\put(45,-6){(c2)}
\end{overpic}
\end{center}
\end{minipage}
\hfill
\begin{minipage}[t]{0.23\linewidth}
\begin{center}
\begin{overpic}[scale=0.09]{tree.png}
		\put(3,27){$P_{c5|c6|c7}^+$}
		
		\put(36,3){$D^{\ast -}$}
		
		\put(35,33){$\Sigma_{c}^{\ast ++}$}
		\put(68,52){$\Lambda_{c}^+$}
		\put(69,20){$\pi^+$}
		\put(45,-6){(d1)}
\end{overpic}
\end{center}
\end{minipage}
\hfill
\begin{minipage}[t]{0.23\linewidth}
\begin{center}
\begin{overpic}[scale=0.09]{tree.png}
		\put(3,27){$P_{c5|c6|c7}^+$}
		
		\put(35,3){$\bar{D}^{\ast 0}$}
		
		\put(38,33){$\Sigma_{c}^{\ast +}$}
		\put(68,52){$\Lambda_{c}^+$}
		\put(69,20){$\pi^0$}
		\put(45,-6){(d2)}
\end{overpic}
\end{center}
\end{minipage}
\end{tabular}
\vspace{7pt}
\caption{Tree-level diagrams for the  decays of $P_{c1}\to \bar{D}\Sigma_c \to \bar{D}\Lambda_{c}\pi$ (a), $P_{c2}\to \bar{D}\Sigma_c^{\ast} \to \bar{D}\Lambda_{c}\pi$ (b), $P_{c3|c4}\to \bar{D}^{\ast}\Sigma_c \to \bar{D}^{\ast}\Lambda_{c}\pi$ (c) and $P_{c5|c6|c7}\to \bar{D}^{\ast}\Sigma_c^{\ast} \to \bar{D}^{\ast}\Lambda_{c}\pi$ (d).}
\label{tree1}
\end{center}
\end{figure}
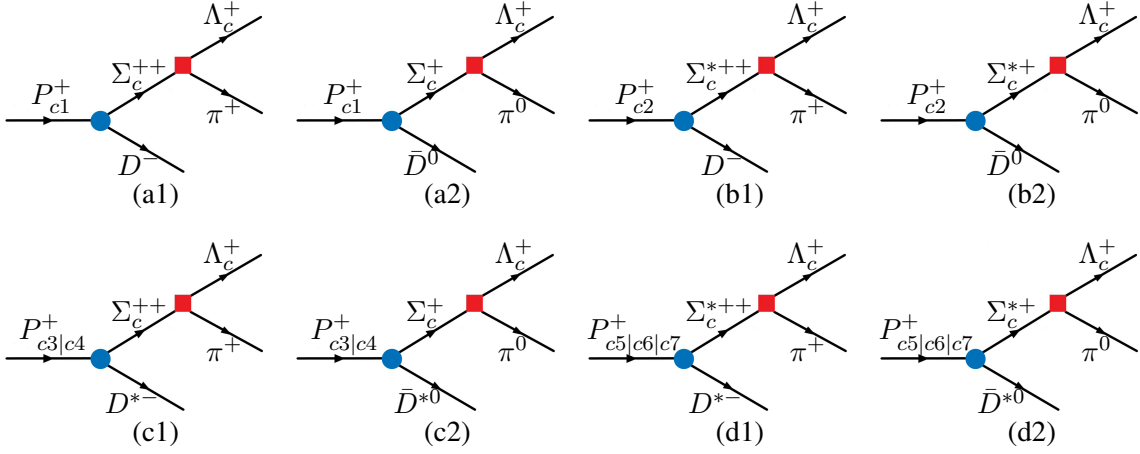

\begin{figure}[ttt]
\begin{center}
\begin{tabular}{ccc}
\begin{minipage}[t]{0.31\linewidth}
\begin{center}
\begin{overpic}[scale=0.26]{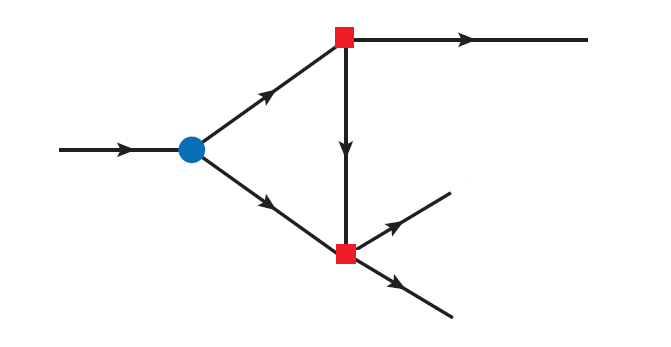}
		\put(11,34){$P_{c3|c4}^+$}
		
		\put(34,14){$\Sigma_{c}^{+}$}
		
		\put(33,39){$\bar{D}^{\ast 0}$}
    	\put(53,28){$D^{-}$}
		\put(67,48){$\pi^+$}
		\put(60,14){$J/\psi(\eta_c)$}
		\put(55,5){$n$}
		\put(45,0){(a1)}
\end{overpic}
\end{center}
\end{minipage}
&
\begin{minipage}[t]{0.31\linewidth}
\begin{center}
\begin{overpic}[scale=0.26]{tr1.png}
		\put(11,34){$P_{c3|c4}^+$}
		
		\put(34,14){$\Sigma_{c}^{+}$}
		
		\put(33,39){$\bar{D}^{\ast 0}$}
    	\put(53,28){$\bar{D}^0$}
		\put(67,48){$\pi^0$}
		\put(60,14){$J/\psi(\eta_c)$}
		\put(55,5){$p$}
		\put(45,0){(a2)}
\end{overpic}
\end{center}
\end{minipage}
&
\begin{minipage}[t]{0.31\linewidth}
\begin{center}
\begin{overpic}[scale=0.26]{tr1.png}
		\put(11,34){$P_{c3|c4}^+$}
		
		\put(32,14){$\Sigma_{c}^{++}$}
		
		\put(33,39){$D^{\ast-}$}
    	\put(53,28){$D^-$}
		\put(67,48){$\pi^0$}
		\put(60,14){$J/\psi(\eta_c)$}
		\put(55,5){$p$}
		\put(45,0){(a3)}
\end{overpic}
\end{center}
\end{minipage}
\\
\begin{minipage}[t]{0.31\linewidth}
\begin{center}
\begin{overpic}[scale=0.26]{tr1.png}
		\put(8,34){$P_{c5|c6|c7}^+$}
		
		\put(33,14){$\Sigma_{c}^{\ast +}$}
		
		\put(33,39){$\bar{D}^{\ast 0}$}
    	\put(53,28){$D^{-}$}
		\put(67,48){$\pi^+$}
		\put(59,13){$J/\psi$}
		\put(55,5){$n$}
		\put(45,0){(b1)}
\end{overpic}
\end{center}
\end{minipage}
&
\begin{minipage}[t]{0.31\linewidth}
\begin{center}
\begin{overpic}[scale=0.26]{tr1.png}
		\put(8,34){$P_{c5|c6|c7}^+$}
		
		\put(33,14){$\Sigma_{c}^{\ast +}$}
		
		\put(33,39){$\bar{D}^{\ast 0}$}
    	\put(53,28){$\bar{D}^0$}
		\put(67,48){$\pi^0$}
		\put(59,13){$J/\psi$}
		\put(55,5){$p$}
		\put(45,0){(b2)}
\end{overpic}
\end{center}
\end{minipage}
&
\begin{minipage}[t]{0.31\linewidth}
\begin{center}
\begin{overpic}[scale=0.26]{tr1.png}
		\put(8,34){$P_{c5|c6|c7}^+$}
		
		\put(30,14){$\Sigma_{c}^{\ast ++}$}
		
		\put(33,39){$D^{\ast-}$}
    	\put(53,28){$D^-$}
		\put(67,48){$\pi^0$}
		\put(59,13){$J/\psi$}
		\put(55,5){$p$}
		\put(45,0){(b3)}
\end{overpic}
\end{center}
\end{minipage}
\end{tabular}
\caption{Triangle-loop diagrams for the  decays of $P_{c3|c4}\to \bar{D}^{\ast}\Sigma_{c} \to J/\psi(\eta_c) N\pi$ (a) and $P_{c5|c6|c7}\to \bar{D}^{\ast}\Sigma_{c}^{\ast} \to J/\psi N\pi$ (b) via the rescattering of final states.}
\label{tree2}
\end{center}
\end{figure}
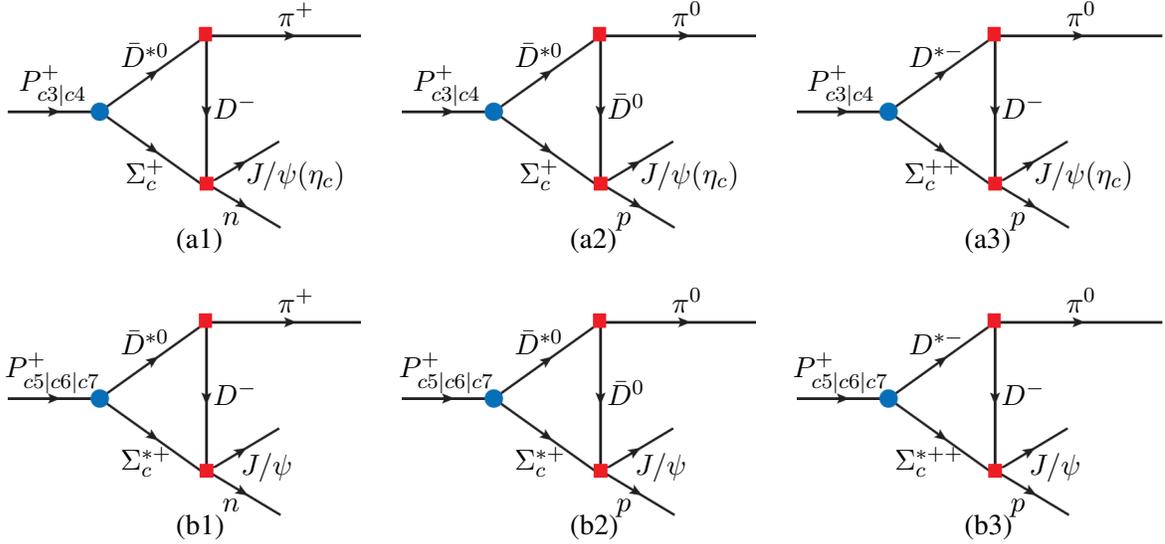

The three-body decay modes of the $\bar{D}^{(\ast)}\Sigma_{c}^{(\ast)}$ molecules can proceed via two types of Feynman diagrams as shown in Fig.~\ref{tree1} and Fig.~\ref{tree2}: tree-level diagrams and triangle-loop diagrams.  The binding energies of the $\bar{D}^{\ast}\Sigma_{c}^{(\ast)}$ molecules relative to their mass thresholds are around 4-20 MeV~\cite{Liu:2019tjn,ParticleDataGroup:2020ssz}, while the mass thresholds of  $\bar{D}\pi\Sigma_{c}^{(\ast)}$ are 5 MeV less than those of $\bar{D}^{\ast}\Sigma_{c}^{(\ast)}$, which implies that the $\bar{D}^{\ast}\Sigma_{c}^{(\ast)}$ molecules decaying into  $\bar{D}\pi\Sigma_{c}^{(\ast)}$ via off-shell $\bar{D}^{\ast}$ meson are heavily suppressed or forbidden. The phase space of $\Sigma_{c}^{(\ast)}\to \Lambda_{c}\pi$ is more than 30 MeV, so that the tree-level decays of $\bar{D}^{(\ast)}\Sigma_{c}^{(\ast)} \to \bar{D}^{(\ast)}\left(\Sigma_{c}^{(\ast)}\to \Lambda_{c}\pi\right) \to \bar{D}^{(\ast)}\Lambda_{c}\pi$ are  allowed,  as shown in Fig.~\ref{tree1}.  The rescattering between final states $\bar{D}^{(\ast)}\Lambda_{c}\pi $ can form the decay modes of the triangle-loop diagrams, which 
are small due to the fact that the loop diagrams are suppressed with respect to the tree-level diagrams~\cite{Yan:2021wdl}, satisfying the power counting of EFT.   Although the decay mode of $\bar{D}^{\ast}\to \bar{D}\pi$ is not allowed at tree level, their final states $\bar{D}\Sigma_{c}^{(\ast)}$ would couple to the $J/\psi N$ and $\eta_{c}N$ channels according to  HQSS~\cite{Sakai:2019qph}.  Therefore, considering the final-state interactions, it would lead to the decay modes via the triangle-loop  mechanism  as shown  in Fig.~\ref{tree2}.

To calculate the Feynman diagrams of Figs.~\ref{tree1} and \ref{tree2}, we have to describe the  interactions related to each vertex, which can be classified into three categories.  The first category involves the interactions between the hadronic molecules and their constituents.  In this work, the seven $\bar{D}^{(\ast)}\Sigma_{c}^{(\ast)}$ hadronic molecules  are denoted by $P_{c1}$, $P_{c2}$, $\ldots$, $P_{c7}$,  following the order of scenario A of Table~I in  Ref.~\cite{Liu:2019tjn} (see also Table II). Here, we should note that $P_{c1}$, $P_{c3}$, and $P_{c5}$ have $J=1/2$, $P_{c2}$ and $P_{c4}$ have $J=3/2$, and $P_{c7}$ has $J=5/2$, respectively.    Their interactions with the corresponding constituents are described by the following Lagrangians~\cite{Lin:2019qiv}:
\begin{eqnarray}
\nonumber
\mathcal{L}_{P_{c1}\bar{D} \Sigma_{c}}&=&-i g_{P_{c1}  \bar{D} \Sigma_{c}} \bar{\Sigma}_{c}    \bar{D} P_{c1} ,  
\\ \nonumber
\mathcal{L}_{P_{c2}\bar{D} \Sigma_{c}^{\ast}}&=&-i g_{P_{c2}\bar{D} \Sigma_{c}^{\ast}} \bar{\Sigma}_{c\mu}^{\ast}  \bar{D} P_{c2}^{\mu} ,   
\\ \nonumber
\mathcal{L}_{P_{c3}\bar{D}^{\ast} \Sigma_{c}}&=&g_{P_{c3}\bar{D}^{\ast} \Sigma_{c}} \bar{\Sigma}_{c} \gamma^{5}  \widetilde{\gamma}^{\mu} \bar{D}^{\ast}_{\mu} P_{c3},  \\ 
 \mathcal{L}_{P_{c4}\bar{D}^{\ast} \Sigma_{c}}&=&g_{P_{c4}\bar{D}^{\ast} \Sigma_{c}}  \bar{\Sigma}_{c}  \bar{D}^{\ast}_{\mu} P_{c4}^{\mu} ,
\\ \nonumber
\mathcal{L}_{P_{c5}\bar{D}^{\ast} \Sigma_{c}^{\ast}}&=&g_{P_{c5}\bar{D}^{\ast} \Sigma_{c}^{\ast}} \bar{\Sigma}_{c}^{\ast\mu}   \bar{D}^{\ast}_{\mu}  P_{c5} ,  
\\ \nonumber
\mathcal{L}_{P_{c6}\bar{D}^{\ast} \Sigma_{c}^{\ast}}&=&g_{P_{c6}\bar{D}^{\ast} \Sigma_{c}^{\ast}} \bar{\Sigma}_{c\mu}^{\ast}  \bar{D}^{\ast}_{\nu}\gamma^{5}\widetilde{\gamma}^{\nu} P_{c6}^{\mu}, 
\\ \nonumber
\mathcal{L}_{P_{c7}\bar{D}^{\ast} \Sigma_{c}^{\ast}}&=&g_{P_{c7}\bar{D}^{\ast} \Sigma_{c}^{\ast}} \bar{\Sigma}_{c\mu}^{\ast} \bar{D}^{\ast}_{\nu} P_{c7}^{\mu\nu} , 
\end{eqnarray}
where  $g_{P_{c1...c7}\bar{D}^{(\ast)} \Sigma_{c}^{(\ast)}}$ are the couplings, which will be determined by the EFT approach described below. $\widetilde{\gamma}^{\mu}$ is defined as  $\widetilde{\gamma}^{\mu}=(g^{\mu\nu}-p^{\mu}p^{\nu}/p^2){\gamma_{\nu}}$, where  $p$ is the momentum of the initial $P_{c}$ states.

The second category involves the decays of the molecular constituents, i.e., $\Sigma_{c}^{(\ast)}\to \Lambda_{c}\pi$ and $\bar{D}^{\ast} \to \bar{D} \pi$.  The Lagrangians describing $\Sigma_{c}^{(\ast)}\to \Lambda_{c}\pi$ read
\begin{eqnarray}
\mathcal{L}_{\pi\Lambda_{c}{\Sigma}_{c}}&=&-i\frac{g_{\pi\Lambda_{c}{\Sigma}_{c}}}{f_{\pi}}~\bar{\Lambda}_{c}\gamma^\mu\gamma_5 \partial_\mu\vec{\phi}_\pi\cdot \vec{\tau}{\Sigma}_{c},\\ \nonumber
\mathcal{L}_{\pi\Lambda_{c}{\Sigma}_{c}^*}&=&\frac{g_{\pi\Lambda_{c}{\Sigma}_{c}^*}}{f_{\pi}}~\bar{\Lambda}_{c}\partial^\mu\vec{\phi}_\pi\cdot \vec{\tau}{\Sigma}_{c\mu}^*,
\end{eqnarray}
where the pion decay constant $f_{\pi}=132$ MeV. The couplings $g_{\pi\Lambda_{c}{\Sigma}_{c}}$ and $g_{\pi\Lambda_{c}{\Sigma}_{c}^*}$  can be determined by reproducing the corresponding experimental data. { With the decay widths of $\Sigma_{c}^{++,+,0}\to \Lambda_{c}^{+}\pi^{+,0,-}$ and $\Sigma_{c}^{\ast ++,+,0}\to \Lambda_{c}^{+}\pi^{+,0,-}$ being 1.89, 2.3, 1.83 MeV and 14.78, 17.2, 15.3 MeV, respectively~\cite{Zyla:2020zbs}, we obtain the corresponding couplings $g_{\pi\Lambda_{c}{\Sigma}_{c}}=0.538,0.554,0.532$ and $g_{\pi\Lambda_{c}{\Sigma}_{c}^*}=0.984,1.043,1.001$,  consistent with other works~\cite{Liu:2011xc,Cheng:2015naa}.}
In addition, we find that the two kinds of  couplings approximately satisfy the relationship $g_{\pi\Lambda_{c}{\Sigma}_{c}^*}=\sqrt{3}g_{\pi\Lambda_{c}{\Sigma}_{c}}$, which can be derived from the quark model~\cite{Liu:2011xc}. 
The Lagrangian describing  the $D^{\ast}$ decay into $D\pi$ is
\begin{eqnarray}
\mathcal{L}_{D D^{\ast} \pi}&=& -i g_{D D^{\ast} \pi} (D \partial^{\mu}\pi D^{\ast\dag}_{\mu}-D_{\mu}^* \partial^{\mu} \pi  D^{\dag}), 
\end{eqnarray}
where the coupling is determined to be $g_{ D^{\ast+}D^{0}\pi^{+} }=16.818$  by  reproducing the decay width of $D^{\ast+}\to D^{0} \pi^+$~\cite{Zyla:2020zbs}.
Experimentally, there exists only  an upper limit $\Gamma<2.1$ MeV for the  $D^{\ast0}$ width.  Thus we turn to the quark model~\cite{Rosner:2013sha}, where the strong decay width  of $D^{\ast0}$ is estimated to be  $\Gamma_{D^{\ast0}\to D^{0} \pi^{0}}=34.658 $ keV.~\footnote{We note that the lattice QCD simulation~\cite{Becirevic:2012pf} gave a relatively larger value,  i.e., $\Gamma_{D^{\ast0}\to D^{0} \pi^{0}}=53\pm9$ keV. From isospin symmetry, we expect that the $D^{*0}$ strong decay width is smaller than the $D^{*+}$ strong decay width because the $D^{*0}\to D^+\pi^-$ decay mode is kinematically forbidden. As a result, we do not adopt the lattice QCD result.}  With these numbers, we obtain the coupling $g_{ D^{\ast0}D^{0}\pi^{0} }=11.688$. It is clear that the strong couplings approximately satisfy  isospin symmetry. 

The last category involves the rescattering  of final states, which is described by the  scattering amplitude $T$ and is responsible for the dynamical generation of the pentaquark states. It can be obtained by solving the   Lippmann-Schwinger equation
\begin{eqnarray}
T=(1-VG)^{-1}V,
\end{eqnarray}
where $V$ is the coupled-channel potential determined by the contact EFT approach described below, and $G$ is the two-body propagator. 
Here to avoid the ultraviolet divergence induced by evaluating the loop function $G$ and keep the unitarity of the $T$ matrix{ \footnote{The  loop function can also be  regularized by other methods such as   momentum cut off scheme and  dimensional regularization scheme~\cite{Oset:1997it,Jido:2003cb,Wu:2010jy,Hyodo:2011ur,Debastiani:2017ewu}.   }}, we include a regulator of Gaussian form { $F(q^2,k)=e^{-2q^2/\Lambda^2}/e^{-2k^2/\Lambda^2}$} in the integral as
\begin{eqnarray}
G(\sqrt{s})= 2m_{1}\int \frac{d^{3}q}{(2\pi)^{3}}\frac{\omega_{1}+\omega_{2}}{2~\omega_{1}\omega_{2}} \frac{F(q^2,k)}{(\sqrt{s})^2-(\omega_1+\omega_2)^2+i \varepsilon}
\label{loopfunction},
\end{eqnarray}
where  { $\sqrt{s}$ }is  the total energy in the center-of-mass (c.m.) frame of $m_{1}$ and $m_{2}$, {   $\omega_{i}=\sqrt{m_{i}^2+q^2}$ is the energy of the particle,}  $\Lambda$ is the momentum cutoff,  and   the c.m. momentum $k$ is ,
\begin{eqnarray}
k=\frac{\sqrt{{s}-(m_{1}+m_{2})^2}\sqrt{s-(m_{1}-m_{2})^2}}{2{\sqrt{s}}}.
\end{eqnarray}

The dynamically generated pentaquark states correspond to poles in the unphysical sheet,   which are below the $\bar{D}^{(\ast)}\Sigma_{c}^{(\ast)}$ mass thresholds   and above the mass thresholds of  $J/\psi N$ and $\eta_{c}N$.  
 The effect of scattering  amplitude  $T$  passing $\sqrt{s}$ to unphysical sheet  has consequences only for the loop functions. We denote the loop function in physical sheet and unphysical sheet  as  
$G_{I}(\sqrt{s})$ and $G_{II}(\sqrt{s})$, respectively, which are related by $G_{II}(\sqrt{s})=G_{I}(\sqrt{s})-2i\mathrm{Im}~G_{I}(\sqrt{s})$~\cite{Oller:1997ti,Roca:2005nm}.   The imaginary part  of Eq.~(\ref{loopfunction})  can be derived via Plemelj-Sokhotski formula, i.g., $\mathrm{Im}~G(\sqrt{s})=-\frac{2m_1 }{8\pi} \frac{k}{\sqrt{s}}$, and then the  Eq.~(\ref{loopfunction})  in the unphysical sheet is written as
\begin{eqnarray}
G_{II}(\sqrt{s})=G_{I}(\sqrt{s}){+~i\frac{2m_1 }{4\pi} \frac{k}{\sqrt{s}}}.
\end{eqnarray}

To study the decays of the $\bar{D}^{(*)}\Sigma_c^{(*)}$ molecules, we first extend our single-channel contact-range EFT~\cite{Liu:2019tjn} to include the $J/\psi N$ and $\eta_c N$ channels.  This can be easily achieved by utilizing HQSS in the following way. First, we express the spin wave function of the $\bar{D}^{(\ast)}\Sigma_{c}^{(\ast)}$ pairs  in terms of the spins of the heavy quarks $s_{1H}$ and $s_{2H}$ and those of the light quark(s) (often referred to as brown 
muck~\cite{Isgur:1991xa,Flynn:1992fm}) $s_{1L}$ and $s_{2L}$, where 1 and 2 denote $\bar{D}^{(*)}$ and $\Sigma_c^{(*)}$, respectively, via the following spin coupling formula,
\begin{eqnarray}
&&|s_{1l}, s_{1h}, j_{1}; s_{2l}, s_{2h},j_{2}; J\rangle =
  \\ \nonumber &&
  \sqrt{(2j_{1}+1)(2j_{2}+1)(2s_{L}+1)(2s_{H}+1)}\left(\begin{matrix}
s_{1l} & s_{2l} & s_{L} \\
s_{1h} & s_{2h} & s_{H} \\
j_{1} & j_{2} & J%
\end{matrix}\right)|s_{1l},
s_{2l}, s_{L}; s_{1h}, s_{2h},s_{H}; J\rangle.
\label{9j}
\end{eqnarray}
More explicitly, for the seven $\bar{D}^{(\ast)}\Sigma_{c}^{(\ast)}$ states, the decompositions read
\begin{eqnarray}
|\Sigma_{c}\bar{D}(1/2^{-})\rangle &=& \frac{1}{2}0_{H}\otimes
{1/2}_{L}+ \frac{1}{2\sqrt{3}}1_{H}\otimes {1/2}_{L}+
\sqrt{\frac{2}{3}}1_{H}\otimes {3/2}_{L},   \\ \nonumber
|\Sigma_{c}^{\ast}\bar{D}(3/2^{-})\rangle &=&
-\frac{1}{2}0_{H}\otimes {3/2}_{L}+ \frac{1}{\sqrt{3}}1_{H}\otimes
{1/2}_{L}+ \frac{\sqrt{\frac{5}{3}}}{2}1_{H}\otimes {3/2}_{L},  \\ \nonumber
|\Sigma_{c}\bar{D}^{\ast}(1/2^{-})\rangle &=&
\frac{1}{2\sqrt{3}}0_{H}\otimes {1/2}_{L}+ \frac{5}{6}1_{H}\otimes
{1/2}_{L}-\frac{\sqrt{2}}{3}1_{H}\otimes {3/2}_{L},  \\ \nonumber
|\Sigma_{c}\bar{D}^{\ast}(3/2^{-})\rangle &=&
\frac{1}{\sqrt{3}}0_{H}\otimes {3/2}_{L}- \frac{1}{3}1_{H}\otimes
{1/2}_{L}+ \frac{\sqrt{5}}{3}1_{H}\otimes {3/2}_{L},   \\ \nonumber
|\Sigma_{c}^{\ast}\bar{D}^{\ast}(1/2^{-})\rangle &=&
\sqrt{\frac{2}{3}}0_{H}\otimes {1/2}_{L}-
\frac{\sqrt{2}}{3}1_{H}\otimes {1/2}_{L}-\frac{1}{3}1_{H}\otimes
{3/2}_{L},                                 \\ \nonumber
|\Sigma_{c}^{\ast}\bar{D}^{\ast}(3/2^{-})\rangle &=&
\frac{\sqrt{\frac{5}{3}}}{2}0_{H}\otimes {3/2}_{L}+
\frac{\sqrt{5}}{3}1_{H}\otimes {1/2}_{L}- \frac{1}{6}1_{H}\otimes
{3/2}_{L},  \\ \nonumber
|\Sigma_{c}^{\ast}\bar{D}^{\ast}(5/2^{-})\rangle &=& 1_{H}\otimes
{3/2}_{L}.
\end{eqnarray}

In the heavy quark limit, the $\bar{D}^{(\ast)}\Sigma_{c}^{(\ast)}$ interactions are independent of the spin of the heavy quark, and therefor the potentials can be parameterized by two coupling constants  describing interactions between light quarks of spin 1/2 and 3/2, respectively, i.e.,  $F_{1/2}=\langle 1/2_{L} | V| 1/2_{L} \rangle$  and $F_{3/2}=\langle 3/2_{L} | V| 3/2_{L} \rangle$:
\begin{eqnarray}
V_{\Sigma_{c}\bar{D}}(1/2^{-}) &=& \frac{1}{3}F_{1/2L}+
\frac{2}{3}F_{3/2L},   \\ \nonumber
V_{\Sigma_{c}^{\ast}\bar{D}}(3/2^{-}) &=&
\frac{1}{3}F_{1/2L}+
\frac{2}{3}F_{3/2L}, \\ \nonumber
V_{\Sigma_{c}\bar{D}^{\ast}}(1/2^{-}) &=&
\frac{7}{9}F_{1/2L}+
\frac{2}{9}F_{3/2L}, \\ \nonumber
V_{\Sigma_{c}\bar{D}^{\ast}}(3/2^{-}) &=&
\frac{1}{9}F_{1/2L}+
\frac{8}{9}F_{3/2L},  \\ \nonumber
V_{\Sigma_{c}^{\ast}\bar{D}^{\ast}}(1/2^{-}) &=&\frac{8}{9}F_{1/2L}+
\frac{1}{9}F_{3/2L},   \\ \nonumber
V_{\Sigma_{c}^{\ast}\bar{D}^{\ast}}(3/2^{-}) &=&\frac{5}{9}F_{1/2L}+
\frac{4}{9}F_{3/2L},  \\ \nonumber
V_{\Sigma_{c}^{\ast}\bar{D}^{\ast}}(5/2^{-}) &=& F_{3/2L},
\end{eqnarray}
which can be rewritten as a combination of $C_{a}$ and $C_{b}$, i.e. $F_{1/2} = C_a-2C_b$ and  $F_{3/2} = C_a+C_b$~\cite{Liu:2019tjn}. For the inelastic potentials, one can see that there are three possible channels, $\eta_{c} N$, $J/\psi N$ and $J/\psi \Delta$. However, the $J/\psi \Delta$ channel is suppressed due to  isospin symmetry breaking. From  HQSS, the potentials between $\bar{D}^{(\ast)}\Sigma_{c}^{(\ast)}$ and $J/\psi N,\eta_{c}N$ are only related to  the spin of the light quark $1/2$, denoted by one coupling:  $g=\langle \bar{D}^{(\ast)}\Sigma_{c}^{(\ast)} | 1_{H}\otimes 1/2_{L} \rangle= \langle \bar{D}^{(\ast)}\Sigma_{c}^{(\ast)} | 0_{H}\otimes 1/2_{L} \rangle $. The  potentials of $J/\psi N \to J/\psi N  $, $J/\psi N  \to \eta_{c}N$ and $\eta_{c} N  \to \eta_{c}N$ are suppressed  due to the Okubo-Zweig-Iizuka (OZI) rule, which is also supported by  lattice QCD simulations~\cite{Skerbis:2018lew}. In this work, we take the potentials of $V_{J/\psi(\eta_{c}) N \to J/\psi(\eta_{c}) N }=0$.     In the following, we explicitly show the potentials for all the seven states.

For the $\eta_{c}N-J/\psi N-\bar{D}\Sigma_{c}$ and $\eta_{c}N-J/\psi N-\bar{D}\Sigma_{c}^{\ast}$ coupled channels, the contact-range potentials $V$ in matrix form read 
\begin{equation}
    V_{\eta_{c}N-J/\psi N-\bar{D}\Sigma_{c}}^{J=1/2}=\begin{pmatrix}0&0&\frac{1}{2}g\\0&0&\frac{1}{2\sqrt{3}}g\\\frac{1}{2}g&\frac{1}{2\sqrt{3}}g&C_a \end{pmatrix}, \quad
V_{\eta_{c}N-J/\psi N-\bar{D}\Sigma_{c}^{\ast}}^{J=3/2}=\begin{pmatrix}0&0&0\\0&0&\frac{1}{\sqrt{3}}g\\0&\frac{1}{\sqrt{3}}g&C_a \end{pmatrix}.
\end{equation}
For the $\eta_{c}N-J/\psi N-\bar{D}^{*}\Sigma_{c}$ coupled channels, the contact-range potentials for $J=1/2$ and $J=3/2$ read
\begin{equation}
    V_{\eta_{c}N-J/\psi N-\bar{D}^{*}\Sigma_{c}}^{J=1/2}=\begin{pmatrix}0&0&\frac{1}{2\sqrt{3}}g\\0&0&\frac{5}{6}g\\\frac{1}{2\sqrt{3}}g&\frac{5}{6}g&C_a-\frac{4}{3}C_{b} \end{pmatrix},\quad V_{\eta_{c}N-J/\psi N-\bar{D}^{*}\Sigma_{c}}^{J=3/2}=\begin{pmatrix}0&0&0\\0&0&-\frac{1}{3}g\\0&-\frac{1}{3}g&C_a+\frac{2}{3}C_{b} \end{pmatrix}.
\end{equation}
Similarly for the $\eta_{c}N-J/\psi N-\bar{D}^{*}\Sigma_{c}^{*}$ coupled channels, they read
\begin{equation}
    V_{\eta_{c}N-J/\psi N-\bar{D}^{*}\Sigma_{c}^{*}}^{J=1/2}=\begin{pmatrix}0&0&\sqrt{\frac{2}{3}}g\\0&0&-\frac{\sqrt{2}}{3}g\\\sqrt{\frac{2}{3}}g&-\frac{\sqrt{2}}{3}g&C_a-\frac{5}{3}C_{b} \end{pmatrix},\quad V_{\eta_{c}N-J/\psi N-\bar{D}^{*}\Sigma_{c}^{*}}^{J=3/2}=\begin{pmatrix}0&0&0\\0&0&\frac{\sqrt{5}}{3}g\\0&\frac{\sqrt{5}}{3}g&C_a-\frac{2}{3}C_{b} \end{pmatrix},
\end{equation}
The $J=5/2$ case contains only one channel , i.e., $\bar{D}^{*}\Sigma_{c}^{*}$ (as we have neglected the $J/\psi \Delta$ channel), and the corresponding $V$ reads
\begin{equation}
    V_{\eta_{c}N-J/\psi N-\bar{D}^{*}\Sigma_{c}^{*}}^{J=5/2}=\begin{pmatrix}0&0&0\\0&0&0\\0&0&C_a+C_{b} \end{pmatrix}.
\end{equation}

Using the potentials above we can search for  poles generated by the coupled-channel interactions, and  determine the 
 couplings between the molecular states and their constituents from the residues of the corresponding poles, 
\begin{eqnarray}
g_{i}g_{j}=\lim_{{\sqrt{s}}\to {\sqrt{s_0}}}\left({\sqrt{s}}-{\sqrt{s_0}}\right)T_{ij}(\sqrt{s}),
\end{eqnarray}
where $g_{i}$ denotes the coupling of channel $i$ to the  dynamically generated state and ${\sqrt{s_0}}$ is the pole position.    

With the above Lagrangians we can obtain the amplitudes of the two decay modes, which are explicitly given in Appendix~\ref{appendix B}.
The partial decay widths of $P_{c} \to \bar{D}^{(\ast)}\Lambda_{c}\pi$ and $P_{c} \to J/\psi(\eta_c) N\pi$ as a function of $m_{12}^2$ and $m_{23}^2$~\cite{Zyla:2020zbs} read
\begin{equation}
d\Gamma =  \frac{1}{(2 \pi)^{3}}\frac{1}{2J+1} \frac{\overline{|\mathcal{M}|^2}}{32 m_{P_{c}}^{3}} d m_{12}^{2} d m_{23}^{2},
\end{equation}
with $m_{12}$ the invariant mass of $\Lambda_c \pi$ or $J/\psi(\eta_{c}) N$, and $m_{23}$ the invariant mass of $\bar{D}^{(\ast)}\pi$ or $J/\psi(\eta_{c}) \pi$ for the $P_{c} \to \bar{D}^{(\ast)}\Lambda_{c}\pi$ or $P_{c} \to J/\psi(\eta_{c}) N\pi$ decay, respectively.

\section{Results and discussions}
\label{results}

\begin{table}[!h]
\caption{Masses and quantum numbers of relevant hadrons used in this work~\cite{ParticleDataGroup:2020ssz}. \label{mass}}
\begin{tabular}{ccc|ccc|ccc}
  \hline\hline
   Hadron & $I (J^P)$ & M (MeV) &    Hadron & $I (J^P)$ & M (MeV) &    Hadron & $I (J^P)$ & M (MeV)     \\
  \hline
        $\Sigma_{c}^{++}$ & $1(1/2^+)$ & $2453.97$ & 
      $\Sigma_{c}^{+}$ & $1(1/2^+)$ & $2452.65$ & 
      $\Sigma_{c}^{0}$ & $1(1/2^+)$ & $2453.75$ \\
  $\Sigma_{c}^{\ast ++}$ & $1(3/2^+)$ & $2518.41$ & 
      $\Sigma_{c}^{\ast+}$ & $1(3/2^+)$ & $2517.4$ & 
      $\Sigma_{c}^{\ast0}$ & $1(3/2^+)$ & $2518.48$ \\
        $\pi^{\pm}$ & $1(0^-)$ & $139.57$ & 
      $\pi^{0}$ & $1(0^-)$ & $134.98$ & 
      $\Lambda_{c}^{+}$ & $0(1/2^+)$ & $2286.46$ \\
   $\bar{D}^{0}$ & $\frac{1}{2}(0^-)$ & $1864.84$  &    $D^{-}$ & $\frac{1}{2}(0^-)$ & $1869.66$  & 
      $p$ & $\frac{1}{2}(1/2^+)$ & $938.27$ \\
  $\bar{D}^{\ast0}$ & $\frac{1}{2}(1^-)$ & $2006.85$ &  $D^{\ast-}$ & $\frac{1}{2}(1^-)$ & $2010.26$   & 
      $n$ & $\frac{1}{2}(1/2^+)$ & $939.57$ \\
  $J/\psi$ & $0(1^-)$ & $3096.90$& 
  $\eta_{c}$ & $0(0^-)$ & $2983.90$ \\
 \hline \hline
\end{tabular}
\label{tab:masses}
\end{table}

  In this work, we assume that the pentaquark states are generated by the $\bar{D}^{(\ast)}\Sigma_{c}^{(\ast)}$, $\eta_{c}N$ and $J/\psi N$ coupled channels, and neglect the $\bar{D}^{(\ast)}\Lambda_{c}$ contribution, which  is shown to be rather small with respect to the other three channels in the chiral unitary approach~\cite{Xiao:2019aya,Xiao:2020frg}.    As a result, there are three unknown parameters in the contact-range potentials. Following our previous work~\cite{Liu:2019tjn}, we study two scenarios A and B. In Scenario A, the spins of $P_{c}(4440)$ and  $P_{c}(4457)$ are $1/2$ and $3/2$, respectively, while in Scenario B they are $3/2$ and $1/2$.  For such  coupled-channel systems,  the mass splittings between these channels can be  up to 600 MeV. Therefore, we choose the value of the cutoff in the Gaussian regulator as $\Lambda=1.5$ GeV~\cite{Peng:2020hql}.~\footnote{We have adopted a larger value  $\Lambda=3$ GeV to check the uncertainties induced by the cutoff, and found that it affects little the pole couplings to $\eta_{c}N$ and $J/\psi N$.   }   We tabulate the masses and quantum numbers of relevant particles in Table~\ref{tab:masses}.  
  
\subsection{Pole position and couplings }

\begin{table}[!h]
\centering
\caption{Couplings of the contact-range potentials (in units of GeV$^{-1}$) for Scenario A and B obtained with a cutoff $\Lambda=1.5$ GeV.
}
\label{couplingconstants}
\begin{tabular}{c c c c c c c c}
  \hline \hline
     Scenario   &~~~~ $\Lambda$ (GeV)   &~~~~ $C_a$  &~~~~   $C_b$   &~~~~   $g_1$    &~~~~  $g_3$   &~~~~  $g_4$ 
         \\ \hline  A   &~~~~ 1.5   &~~~~ -52.750   &~~~~   5.625 &~~~   7.650      &~~~   6.760  &~~~  12.350
         \\   B   &~~~~ 1.5   &~~~~ -56.447    &~~~~   -5.480   &~~~   7.350   &~~~   4.610   &~~~  18.000
         \\
  \hline \hline
\end{tabular}
\end{table}

  The three unknown parameters, $C_{a}$, $C_{b}$,  and $g$,  can in principle be determined by reproducing the masses and widths of $P_{c}(4440)$ and $P_{c}(4457)$. However, we find that with only the two inelastic channels ($J/\psi N$ and $\eta_{c}N$),  we cannot obtain a satisfactory fit of the two decay widths with a single $g$, as already noted in Ref.~\cite{Du:2021fmf}. Therefore,  we finetune  $g$ for each of the $P_c(4312)$, $P_c(4440)$, and $P_c(4457)$ states, and the couplings of  $g$ are denoted by $g_{1}$, ..., $g_{7}$, following the order of $P_{c1}$, ..., $P_{c7}$. For the state $P_{c2}$, we did not try to reproduce that of the $P_c(4380)$ state of the LHCb Collaboration~\cite{Aaij:2015tga}, because quite likely they are not the same state. In Ref.~\cite{Du:2019pij}, a reanalysis of the 2019 LHCb data yields a state that corresponds to our $P_{c2}$, whose width is only about 20 MeV.  In the present work, we assume { $g_2=g_1$}. As shown later this leads to a total decay width for $P_{c2}$ ranging from { 9 MeV to 14 MeV, which is equal to the sum of the three-body partial decay widths ($P_{c2}\to {D^{\ast-}} \Lambda_c^+ \pi^+$ and $P_{c2}\to \bar{D}^{\ast0} \Lambda_c^+ \pi^0$)  and the two-body partial width ($P_{c2}\to J/\psi p$), corresponding to  the results obtained in Scenarios B and A, respectively. We note that the total width  is in reasonable agreement with that of Ref.~\cite{Du:2021fmf}}. In Table~\ref{couplingconstants}, we present the values of $C_{a}$, $C_{b}$, and $g$ ($g_{1}$, $g_{3}$, and $g_{4}$) in Scenario A and B by reproducing the masses and  widths of  $P_c(4440)$, and $P_c(4457)$ and the width of $P_{c}(4312)$. One can see that the value of $g_{1}$    is  relatively close to the value of $g_{3}$, but is much smaller than the value of $g_{4}$, which indicates that
   HQSS for the inelastic channels are heavily broken~\footnote{The other channels such as $P-$wave $\bar{D}\Lambda_{c1}$ and $\bar{D}\Sigma_{c}^{\ast}$ can also contribute to the decay widths of $P_{c4}$~\cite{Burns:2019iih,Peng:2020gwk,Burns:2021jlu}, which could be partially responsible for the large value of $g_{4}$.      }. { Since the mass of $P_{c2}$ is close to $P_{c1}$ and the masses of $P_{c5}$, $P_{c6}$, and $P_{c7}$ are close to $P_{c3}$,    the unknown values of $g_{2}$ and $g_{5}$, $g_{6}$, $g_{7}$   in the present work  are taken to be the same as  $g_{1}$ and $g_{3}$, respectively. }

\begin{table}[!h]
\centering

\caption{Masses of the $\bar{D}^{(\ast)}\Sigma_{c}^{(\ast)}$ molecules as well as their couplings to $\bar{D}^{(\ast)}\Sigma_{c}^{(\ast)}$, $J/\psi N$ and $\eta_c N$ in Scenario A for a cutoff $\Lambda=1.5$ GeV. The bold figures are the experimental values used as inputs~\cite{LHCb:2019kea}.  The symbol - denotes that the  $P_{c}$ state does not couple to that particular channel.  
}
\label{tab:coupling1}
\begin{tabular}{c c c c c c c}
  \hline \hline
     State   &~~~~ Molecule   &~~~~ $J^P$   &~~~~   Mass (MeV)   &~~~~   $g_{P_{c}\bar{D}^{(\ast)}\Sigma_{c}^{(\ast)}}$  &~~~~  $g_{P_{c} J/\psi N}$  &~~~~  $g_{P_{c} \eta_c N}$
         \\ \hline  $P_{c1}$   &~~~~ $\bar{D}\Sigma_c$   &~~~~ $(1/2)^-$   &~~~~   4309.3+\textbf{4.9}$i$  &~~~   2.16     &~~~   0.31    &~~~  0.53
         \\   {$P_{c2}$}   &~~~~ {$\bar{D}\Sigma_c^{\ast}$}   &~~~~ {$(3/2)^-$}   &~~~~   4372.2+4.8$i$   &~~~   2.19   &~~~ 0.62   &~~~   - 
         \\   $P_{c3}$   &~~~~ $\bar{D}^{\ast}\Sigma_c$   &~~~~ $(1/2)^-$   &~~~~  \bf{4440.3}+\bf{10.3}$i$   &~~~   2.60  &~~~   0.83     &~~~  0.29
         \\   $P_{c4}$   &~~~~ $\bar{D}^{\ast}\Sigma_c$   &~~~~ $(3/2)^-$   &~~~~   \bf{4457.3}+\bf{3.2}$i$   &~~~   1.70   &~~~   0.49  &~~~   - 
        \\    {$P_{c5}$}   &~~~~ {$\bar{D}^{\ast}\Sigma_c^{\ast}$}   &~~~~ {$(1/2)^-$}    &~~~~   4502.7+14.0$i$    &~~~   2.68   &~~~   0.48   &~~~  0.83
         \\   {$P_{c6}$}   &~~~~ {$\bar{D}^{\ast}\Sigma_c^{\ast}$}   &~~~~ {$(3/2)^-$}   &~~~~   4510.5+7.2$i$   &~~~   2.31  &~~~   0.71   &~~~  -
         \\   $P_{c7}$   &~~~~ $\bar{D}^{\ast}\Sigma_c^{\ast}$   &~~~~ $(5/2)^-$   &~~~~   4522.8    &~~~   1.49    &~~~   -   &~~~   -
         \\
  \hline \hline
\end{tabular}
\end{table}

\begin{table}[!h]
\centering
\caption{Masses of the $\bar{D}^{(\ast)}\Sigma_{c}^{(\ast)}$ molecules as well as their couplings to $\bar{D}^{(\ast)}\Sigma_{c}^{(\ast)}$, $J/\psi N$ and $\eta_c N$ in Scenario B for a cutoff $\Lambda=1.5$ GeV. The bold figures  are the experimental values used as inputs~\cite{LHCb:2019kea}.  The symbol - denotes that the $P_{c}$ state does not couple to that particular channel.    
}
\label{tab:coupling2}
\begin{tabular}{c c c c c c c}
   \hline \hline
     State   &~~~~ Molecule   &~~~~ $J^P$   &~~~~   Mass (MeV)   &~~~~   $g_{P_{c}\bar{D}^{(\ast)}\Sigma_{c}^{(\ast)}}$  &~~~~  $g_{P_{c} J/\psi N}$  &~~~~  $g_{P_{c} \eta_c N}$
         \\ \hline  $P_{c1}$   &~~~~ $\bar{D}\Sigma_c$   &~~~~ $(1/2)^-$   &~~~~   4303.1+\textbf{4.9}$i$    &~~~   2.41    &~~~   0.31   &~~~   0.54
         \\ {$P_{c2}$}   &~~~~ {$\bar{D}\Sigma_c^{\ast}$}   &~~~~ {$(3/2)^-$}    &~~~~   4366.3+3.2$i$   &~~~   2.41    &~~~   0.50   &~~~  -
         \\   $P_{c3}$   &~~~~ $\bar{D}^{\ast}\Sigma_c$   &~~~~ $(1/2)^-$   &~~~~   \bf{4457.3}+\bf{3.2}$i$   &~~~   1.71    &~~~   0.46  &~~~   0.16
         \\   $P_{c4}$   &~~~~ $\bar{D}^{\ast}\Sigma_c$   &~~~~ $(3/2)^-$   &~~~~   \bf{4440.3}+\bf{10.3}$i$   &~~~   2.60   &~~~   0.88 &~~~   -
         \\  $P_{c5}$   &~~~~ $\bar{D}^{\ast}\Sigma_c^{\ast}$   &~~~~ $(1/2)^-$   &~~~~   4523.7+3.7$i$    &~~~   1.56   &~~~   0.25   &~~~  0.43
        \\   {$P_{c6}$}   &~~~~ {$\bar{D}^{\ast}\Sigma_c^{\ast}$}   &~~~~ {$(3/2)^-$}   &~~~~   4515.9+3.0$i$    &~~~   1.99  &~~~   0.45    &~~~ -
         \\   $P_{c7}$   &~~~~ $\bar{D}^{\ast}\Sigma_c^{\ast}$   &~~~~ $(5/2)^-$   &~~~~   4501.3   &~~~   2.59   &~~~   -   &~~~   -
         \\
  \hline \hline
\end{tabular}
\end{table}

With the parameters in Table~\ref{couplingconstants}, we  search for  poles corresponding to the seven $\bar{D}^{(\ast)}\Sigma_{c}^{(\ast)}$ molecules and determine their couplings to the relevant channels. 
 We present the results obtained in Scenario A and B in Tables~\ref{tab:coupling1} and ~\ref{tab:coupling2}, respectively. Compared with the mass spectra of Ref.~\cite{Liu:2019tjn},  the real part of all the poles remains similar, which reconfirms that  the $\bar{D}^{(\ast)}\Sigma_{c}^{(\ast)}$ channels play the dominant role in dynamically generating the complete HQSS muiltplet of hadronic molecules. Furthermore, the  couplings of the seven states to  $\bar{D}^{(\ast)}\Sigma_{c}^{(\ast)}$  are consistent with those of Ref.~\cite{Sakai:2019qph}. { Note that the couplings $g_{P_{c}\bar{D}^{(\ast)}\Sigma_{c}^{(\ast)}}$ obtained in Table~\ref{couplingconstants} are in isospin basis. From them and assuming isospin symmetry we can derive the couplings in particle basis, e.g.,   $g_{P_{c}{D}^{-(\ast)}\Sigma_{c}^{++(\ast)}}=\sqrt{\frac{2}{3}}g_{P_{c}\bar{D}^{(\ast)}\Sigma_{c}^{(\ast)}}$ and  $g_{P_{c}\bar{D}^{0(\ast)}\Sigma_{c}^{+(\ast)}}=\sqrt{\frac{1}{3}}g_{P_{c}\bar{D}^{(\ast)}\Sigma_{c}^{(\ast)}}$.    }    With the pole positions and couplings to their constituents determined, we can now study the three-body decay widths of the pentaquark states.

\begin{table}[ttt]
\centering
\caption{Partial decay widths (in units of MeV) of $P_c^+ \to {D^{(\ast)-}} \Lambda_c^+ \pi^+$ and $P_c^+ \to \bar{D}^{(\ast)0} \Lambda_c^+ \pi^0$  in Scenario A and Scenario B for  a cutoff $\Lambda=1.5$ GeV.
}
\label{tab:width1}
\begin{tabular}{c c c c c}
  \hline \hline
      Scenario  &~~~~   A   &~~~~    A    &~~~~   B   &~~~~    B
         \\   Mode   &~~~~  ${D^{(\ast)-}} \Lambda_c^+ \pi^+$ &~~~~ $\bar{D}^{(\ast)0} \Lambda_c^+ \pi^0$ &~~~~  ${D^{(\ast)-}} \Lambda_c^+ \pi^+$ &~~~~ $\bar{D}^{(\ast)0} \Lambda_c^+ \pi^0$
         \\ \hline $P_{c1}$    &~~~~  0.034 &~~~~ 0.141  &~~~~  0.004 &~~~~ 0.037
         \\  $P_{c2}$    &~~~~  2.085 &~~~~ 2.166  &~~~~ 1.468 &~~~~ 1.479 
         \\  $P_{c3}$    &~~~~  0.002 &~~~~ 0.033  &~~~~  0.517 &~~~~ 1.793
         \\  $P_{c4}$    &~~~~  0.170 &~~~~ 0.591  &~~~~  0.001 &~~~~ 0.011
         \\  $P_{c5}$    &~~~~  2.508 &~~~~ 2.219  &~~~~  6.906 &~~~~ 6.280
         \\  $P_{c6}$    &~~~~  3.087 &~~~~ 2.758   &~~~~  3.866 &~~~~ 3.529
         \\  $P_{c7}$    &~~~~  2.807 &~~~~ 2.578   &~~~~  1.033 &~~~~ 0.915 \\
  \hline \hline
\end{tabular}
\end{table}

\subsection{Decay widths of tree-level diagrams }

For the decay modes of Fig.~\ref{tree1},  
we present  the corresponding partial decay widths  in Table~\ref{tab:width1}. 
{   The decay width of $P_{c}(4312)$  into $\bar{D}\Lambda_{c}\pi$ is about hundreds and tens of keV in scenarios A and B, respectively,  which accounts for $0.4\%-2\%$ of its total width.  } The decay width of $P_{c}(4457)$  into $\bar{D}\Lambda_{c}\pi$ is up to several MeV, while for $P_{c}(4440)$  it is only tens of keV.  The partial width of $P_{c}(4457)$   accounts for tens of percent of its total width, while for $P_{c}(4440)$ it accounts for less than one percent, which  indicates that the $\bar{D}\Lambda_{c}\pi$  mode is a good channel to detect $P_{c}(4457)$ and verify the molecular nature of $P_{c}(4457)$. In addition, we predict the partial decay widths of the other four molecules  into $\bar{D}^{(\ast)}\Lambda_{c}\pi$, which are about several MeV, some of them are even up to tens of MeV, in agreement with the results of Ref.~\cite{Lin:2019qiv}. Therefore the $P_{c2,5,6,7}$ states  can be observed in the $\bar{D}^{\ast}\Lambda_c\pi$ final states in future experiments, especially $P_{c7}$, which is particularly difficult to be detected in the $J/\psi p$ invariant  mass distribution where only the $D-$wave $J/\psi p$ is allowed. One should note that the partial  decay widths of  the seven   $\bar{D}^{(\ast)}\Sigma_{c}^{(\ast)}$ molecules  into $\bar{D}^{(\ast)}\Lambda_{c}\pi$ in Scenario A and B are
of the same order of magnitude, which indicates that the tree-level decay modes cannot discriminate the spins of $P_{c}(4440)$ and $P_{c}(4457)$.

\subsection{Decay widths of triangle-loop diagrams  }

\begin{figure}[!h]
\centering
\subfigure[\ $P_{c3}$-Scenario A]{
\includegraphics[width=7.6cm]{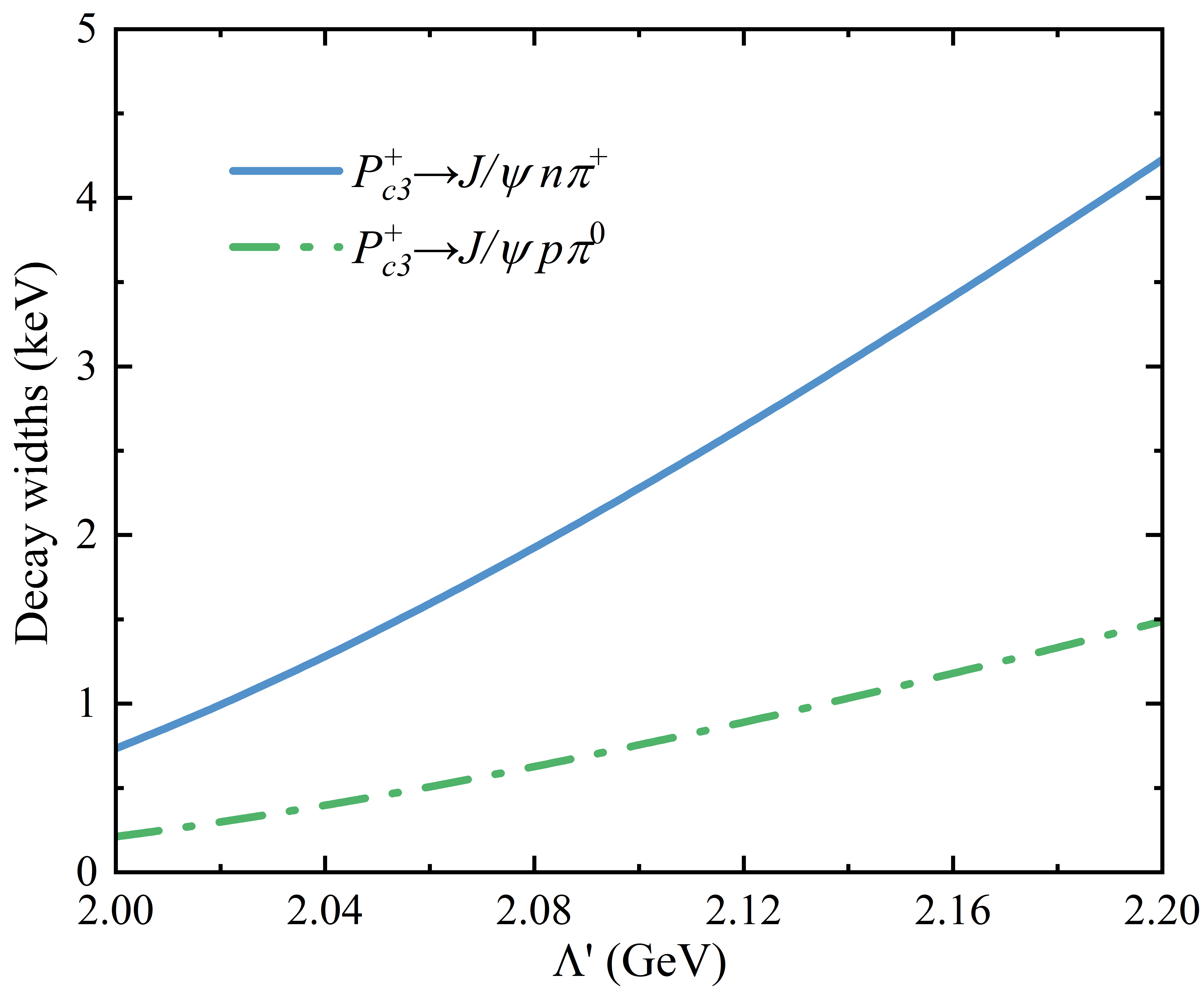}
}
\quad
\subfigure[\ $P_{c3}$-Scenario B]{
\includegraphics[width=7.6cm]{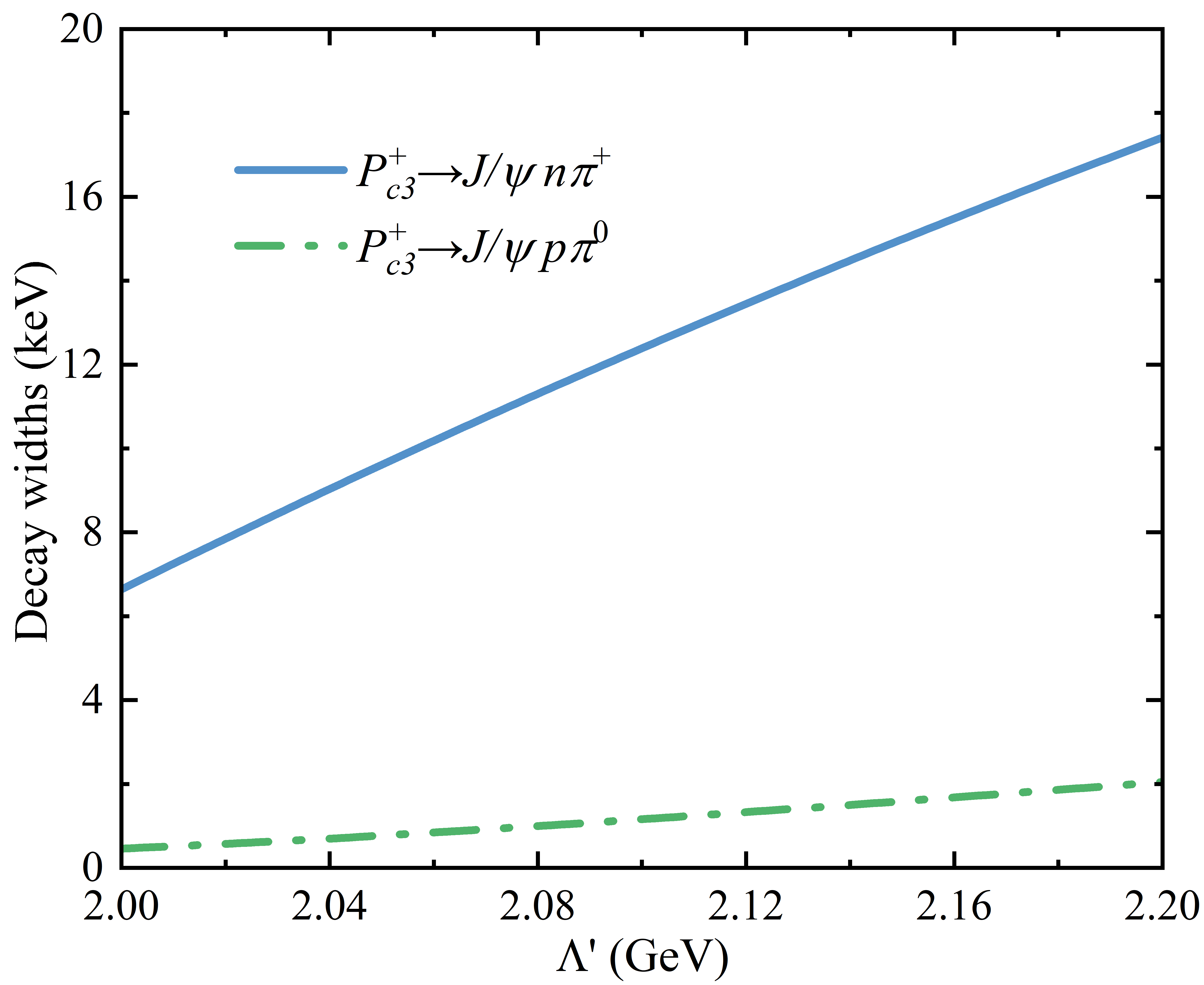}
}
\quad
\subfigure[\ $P_{c4}$-Scenario A]{
\includegraphics[width=7.6cm]{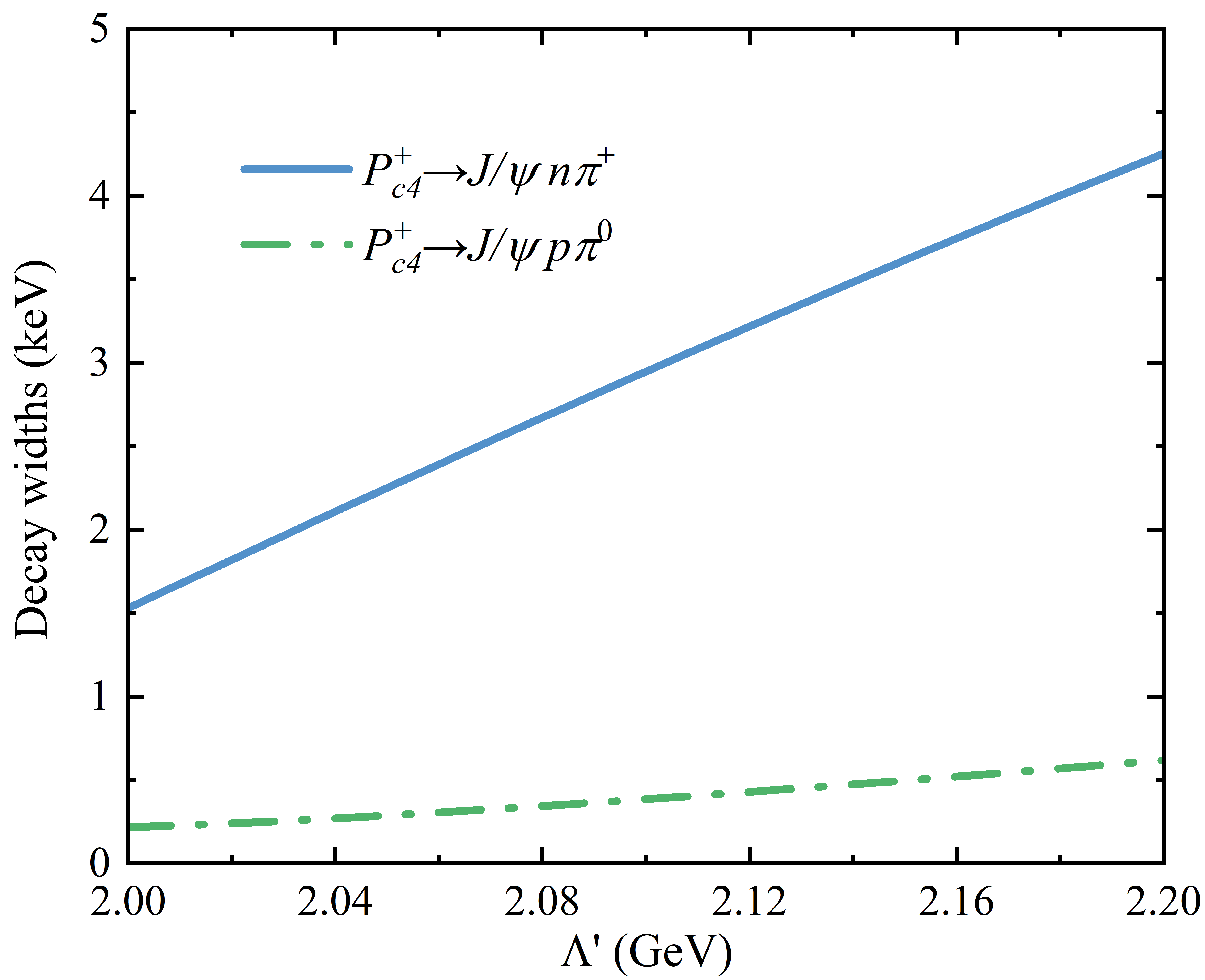}
}
\quad
\subfigure[\ $P_{c4}$-Scenario B]{
\includegraphics[width=7.6cm]{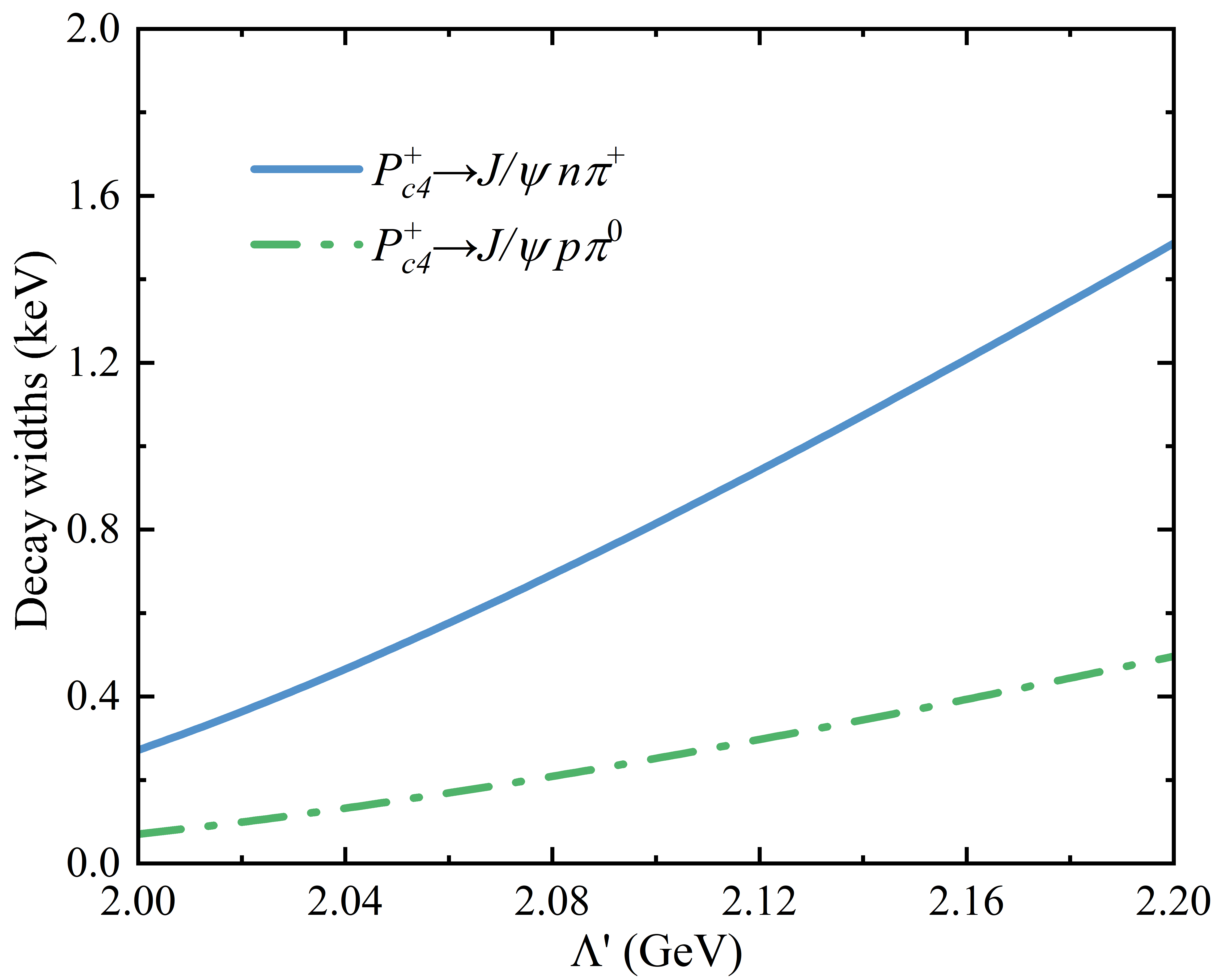}
}
\caption{Partial decay widths of  $P_{c3|c4}^+ \to J/\psi n \pi^+$ and $P_{c3|c4}^+ \to J/\psi p \pi^0$ as a function of the  cutoff $\Lambda'$ in the form factors (\ref{form factor}).  The results  are represented  by the blue solid lines and green dot-dashed lines, respectively, which correspond to the couplings determined for a cutoff  $\Lambda=$1.5 GeV. } 
\label{Pc34}
\end{figure}

    For the triangle-loop diagrams, we have introduced a form factor in Eq.~(\ref{form factor}) to suppress the ultraviolate divergence. The parameter  $\Lambda^{\prime}$ is usually estimated by the relationship  $\Lambda^{\prime}=m_{E}+\alpha \Lambda_{QCD}$, where $m_{E}$  is the mass of the exchanged particle, $\Lambda_{QCD}\sim200-300$~MeV is the scale parameter of quantum chromodynamics (QCD), and $\alpha$ is a dimensionless  parameter of {order} unity. Following our previous work, we vary $\alpha$ from 0.5 to 1.5~\cite{Ling:2021lmq}, and then  $\Lambda^{\prime}$ varies from 2.0 to 2.2 GeV. 
    
 \begin{figure}[!h]
\centering
\subfigure[\ $P_{c3}$-Scenario A]{
\includegraphics[width=7.6cm]{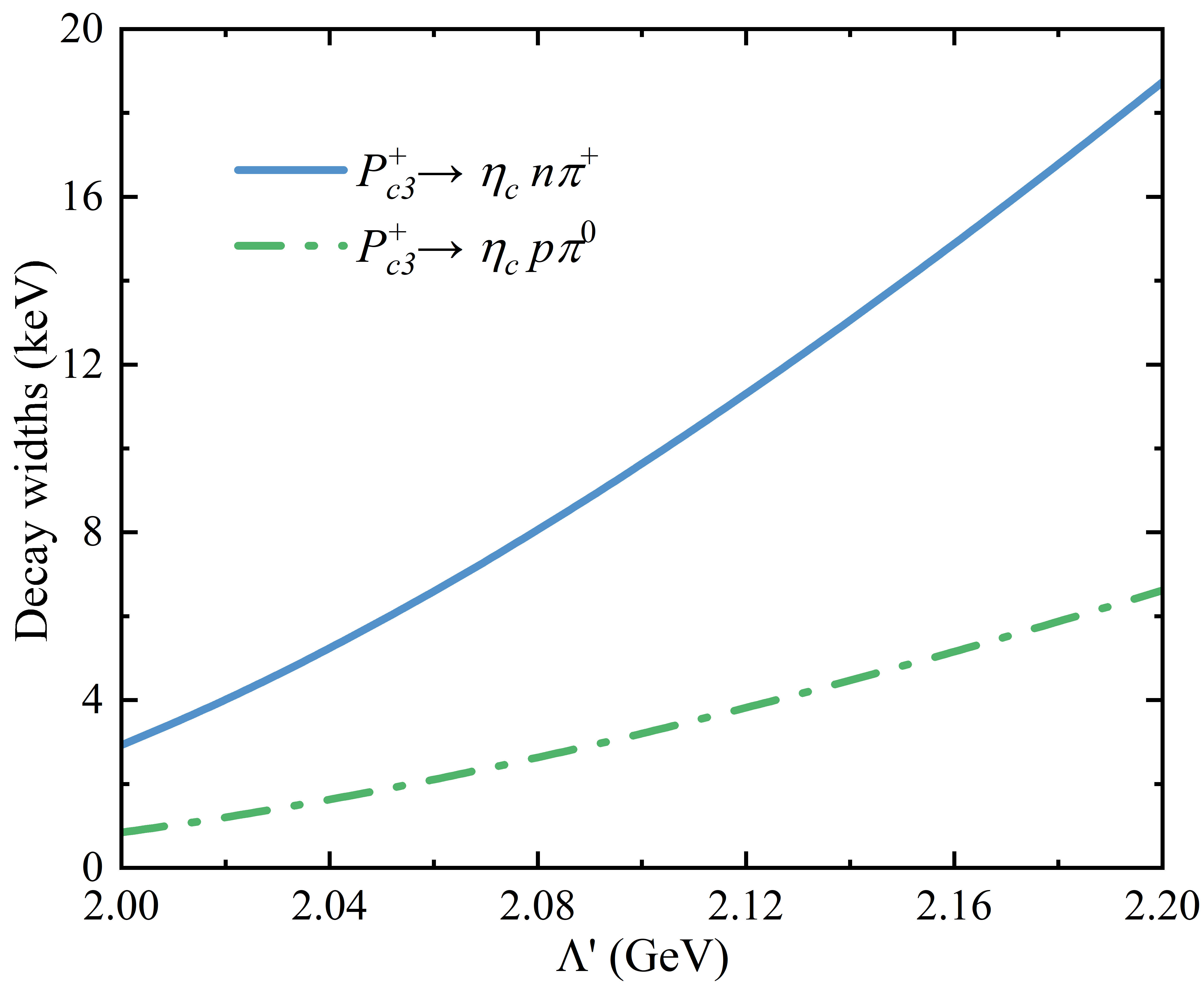}
}
\quad
\subfigure[\ $P_{c3}$-Scenario B]{
\includegraphics[width=7.6cm]{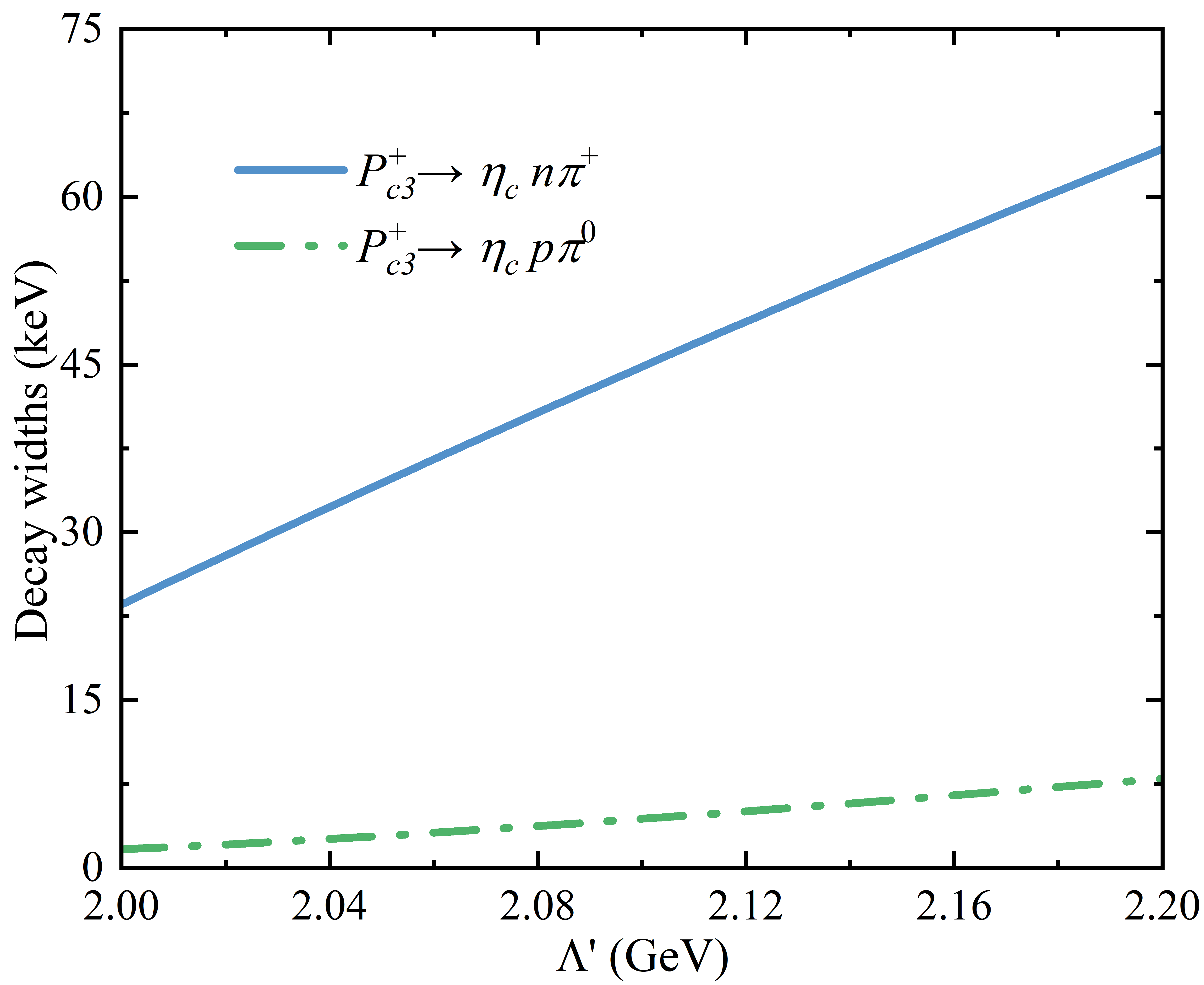}
}
\quad
\subfigure[\ $P_{c4}$-Scenario A]{
\includegraphics[width=7.6cm]{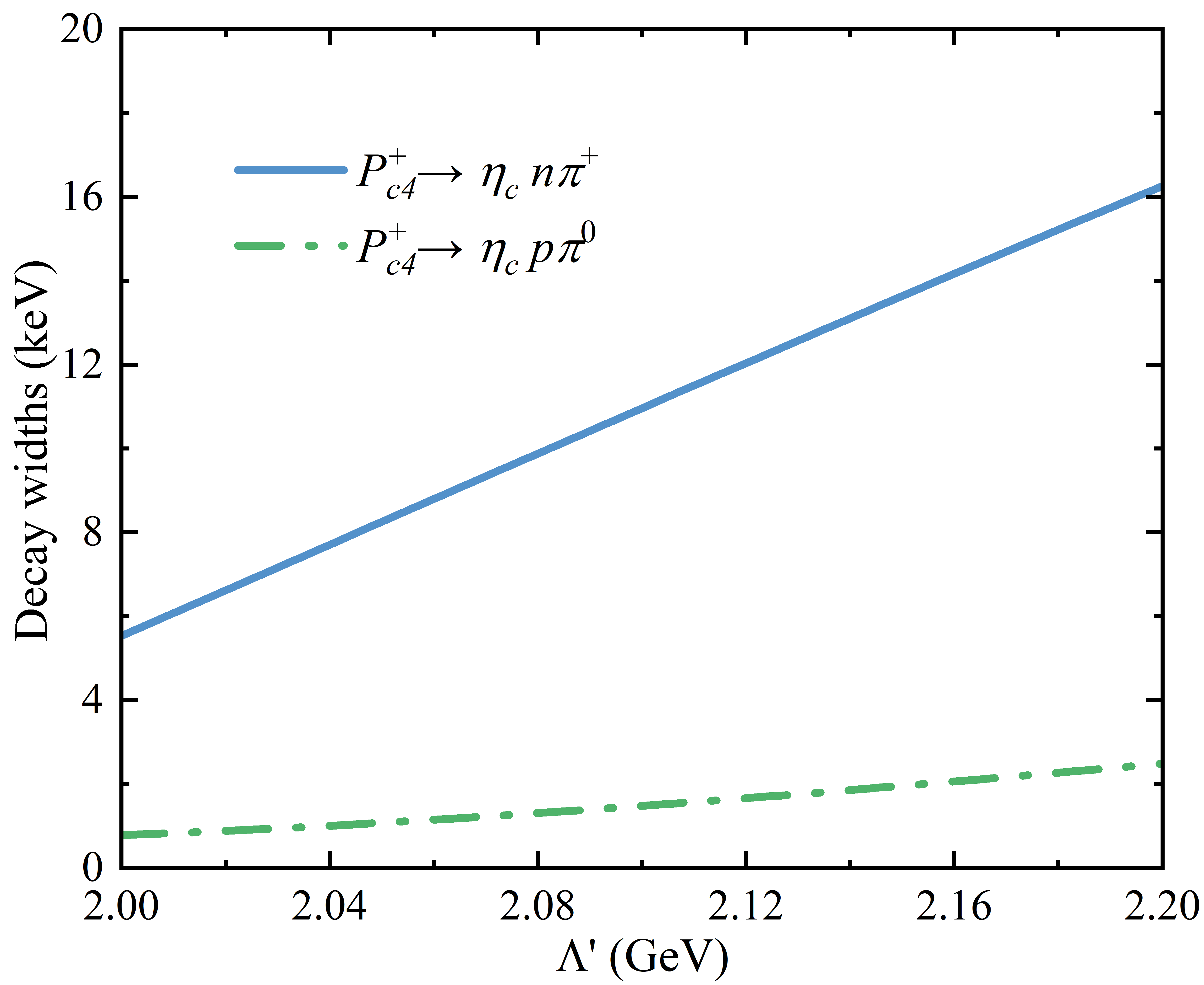}
}
\quad
\subfigure[\ $P_{c4}$-Scenario B]{
\includegraphics[width=7.6cm]{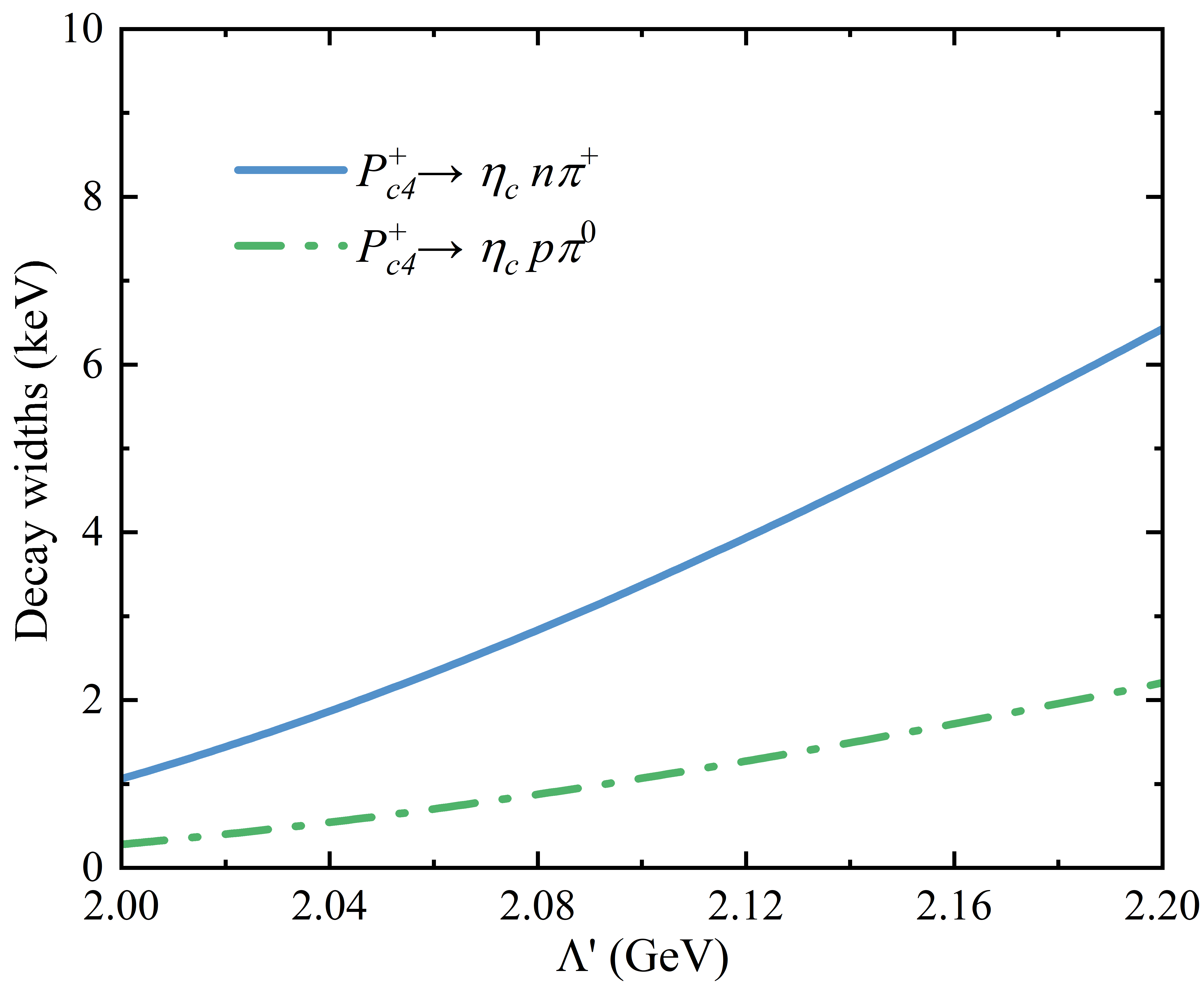}
}
\caption{Partial decay widths of  $P_{c3|c4}^+ \to \eta_{c} n \pi^+$ and $P_{c3|c4}^+ \to \eta_{c} p \pi^0$ as a function of the cutoff $\Lambda'$ in the form factors (\ref{form factor}).    The results  are represented  by the blue solid lines and green dot-dashed lines, respectively, which correspond to the couplings determined for a cutoff  $\Lambda=$1.5 GeV. } 
\label{Pc34-eta}
\end{figure}

In Fig.~\ref{Pc34}, we plot the partial decay widths of $P_{c}(4440)$ and $P_{c}(4457)$ into $J/\psi N \pi$ as a function of the cutoff parameter $\Lambda'$.   One can see that the widths are only several keV in both Scenario A and  B, which are weakly dependent on the variations of both the couplings and the cutoff $\Lambda'$. The widths in both scenarios are similar,  which indicates that  one cannot discriminate the spins of $P_{c}(4440)$ and $P_{c}(4457)$ in the $J/\psi N \pi$ channel, consistent with our previous work~\cite{Ling:2021lmq}.   The partial decay width  of  $P_{c}(4457)$  into $J/\psi N \pi$ account for only 0.1 percent of its total decay width, which  is much smaller than that decaying into $\bar{D}\Lambda_{c}\pi$.

\begin{figure}[htbp]
\centering
\subfigure[\ $P_{c5}$-Scenario A]{
\includegraphics[width=7.2cm]{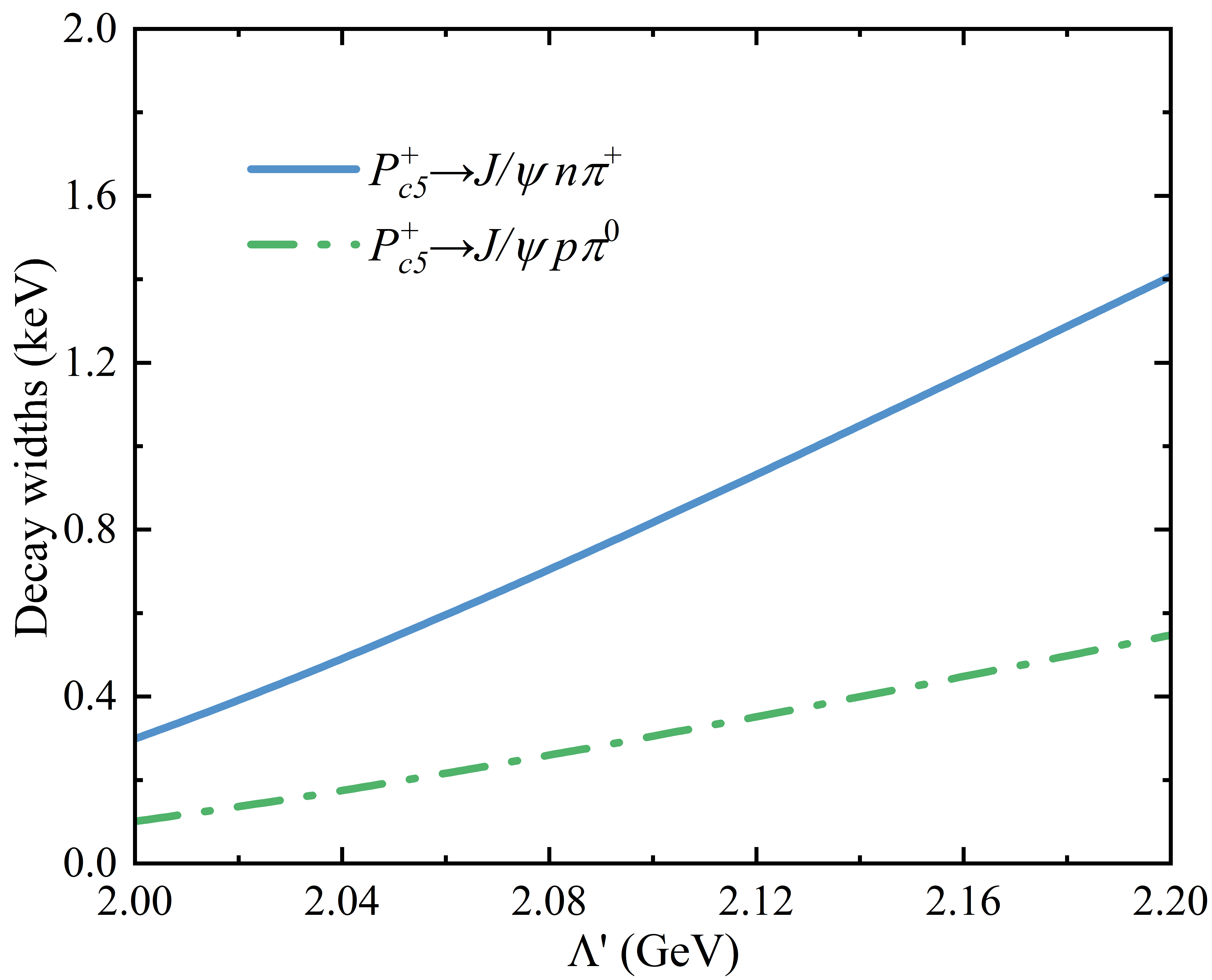}
}
\quad
\subfigure[\ $P_{c5}$-Scenario B]{
\includegraphics[width=7.2cm]{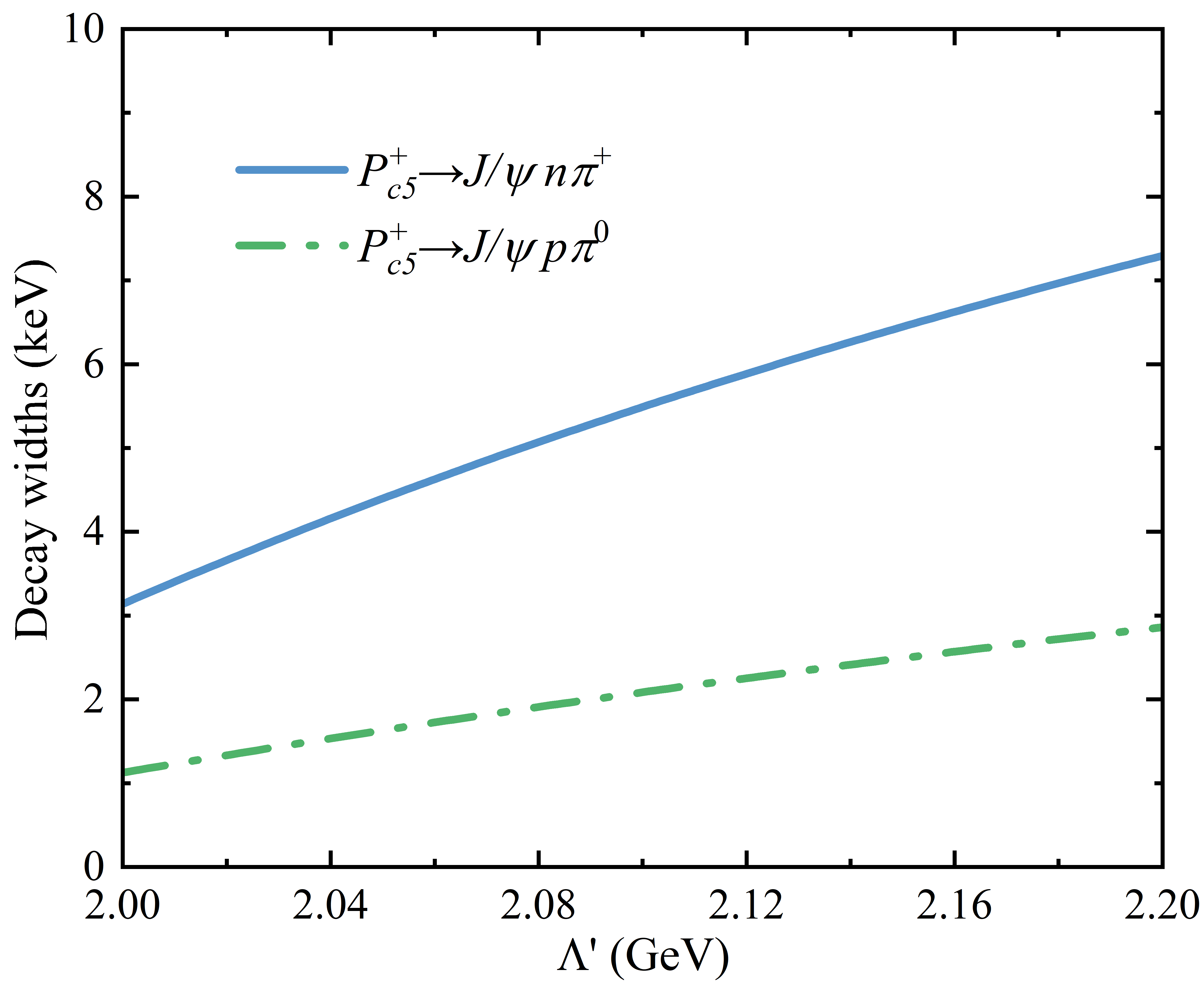}
}
\quad
\subfigure[\ $P_{c6}$-Scenario A]{
\includegraphics[width=7.2cm]{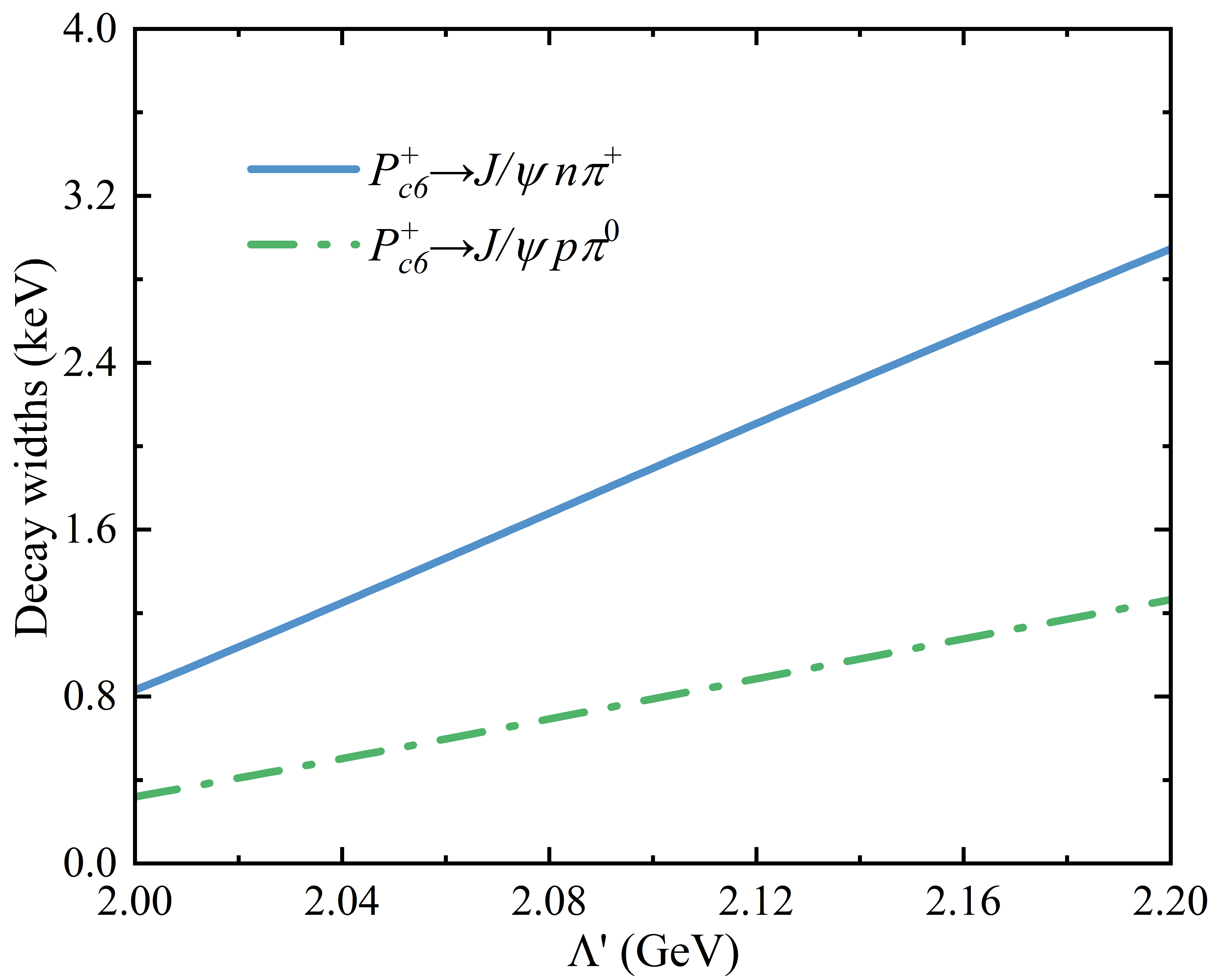}
}
\quad
\subfigure[\ $P_{c6}$-Scenario B]{
\includegraphics[width=7.2cm]{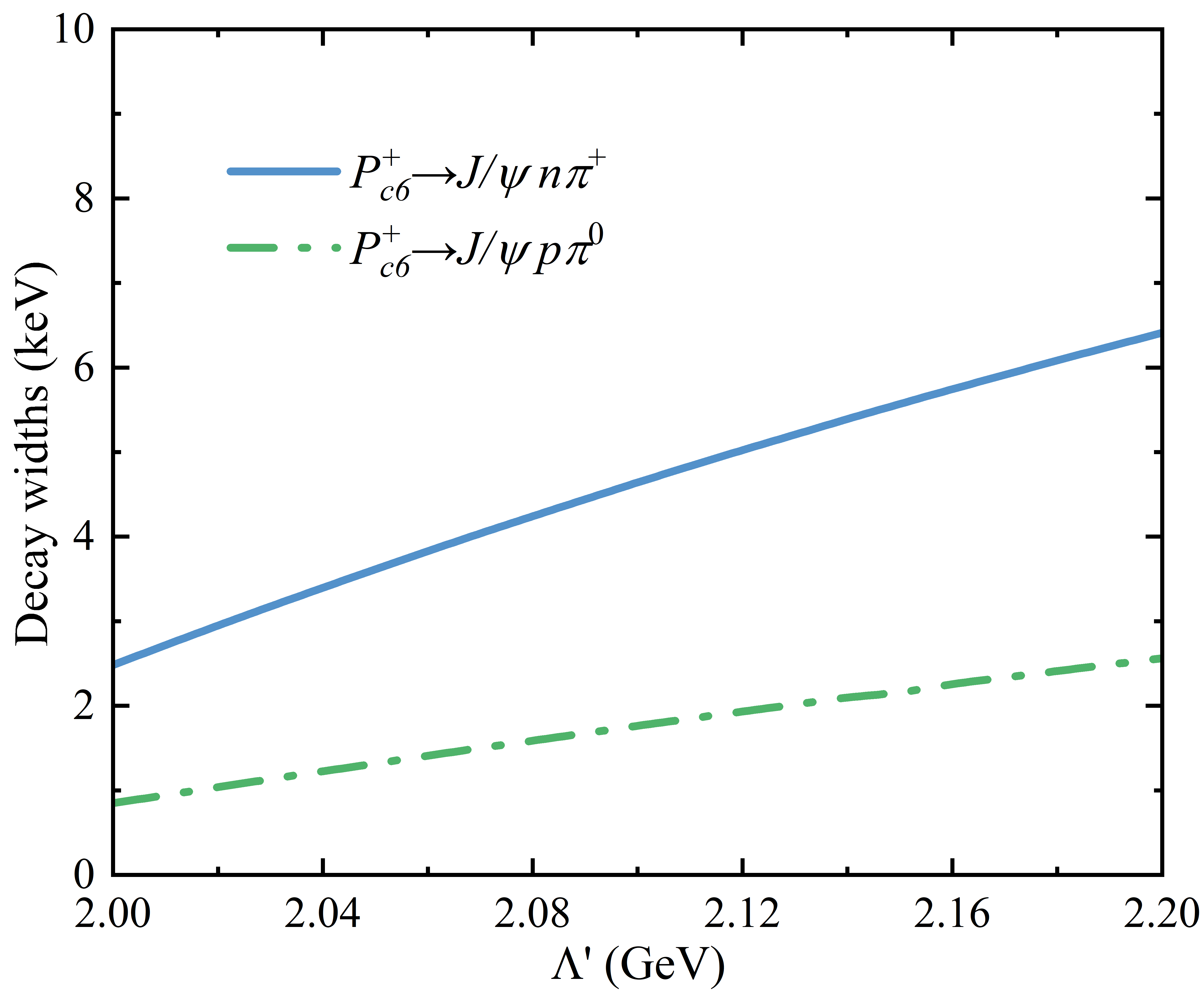}
}
\quad
\subfigure[\ $P_{c7}$-Scenario A]{
\includegraphics[width=7.2cm]{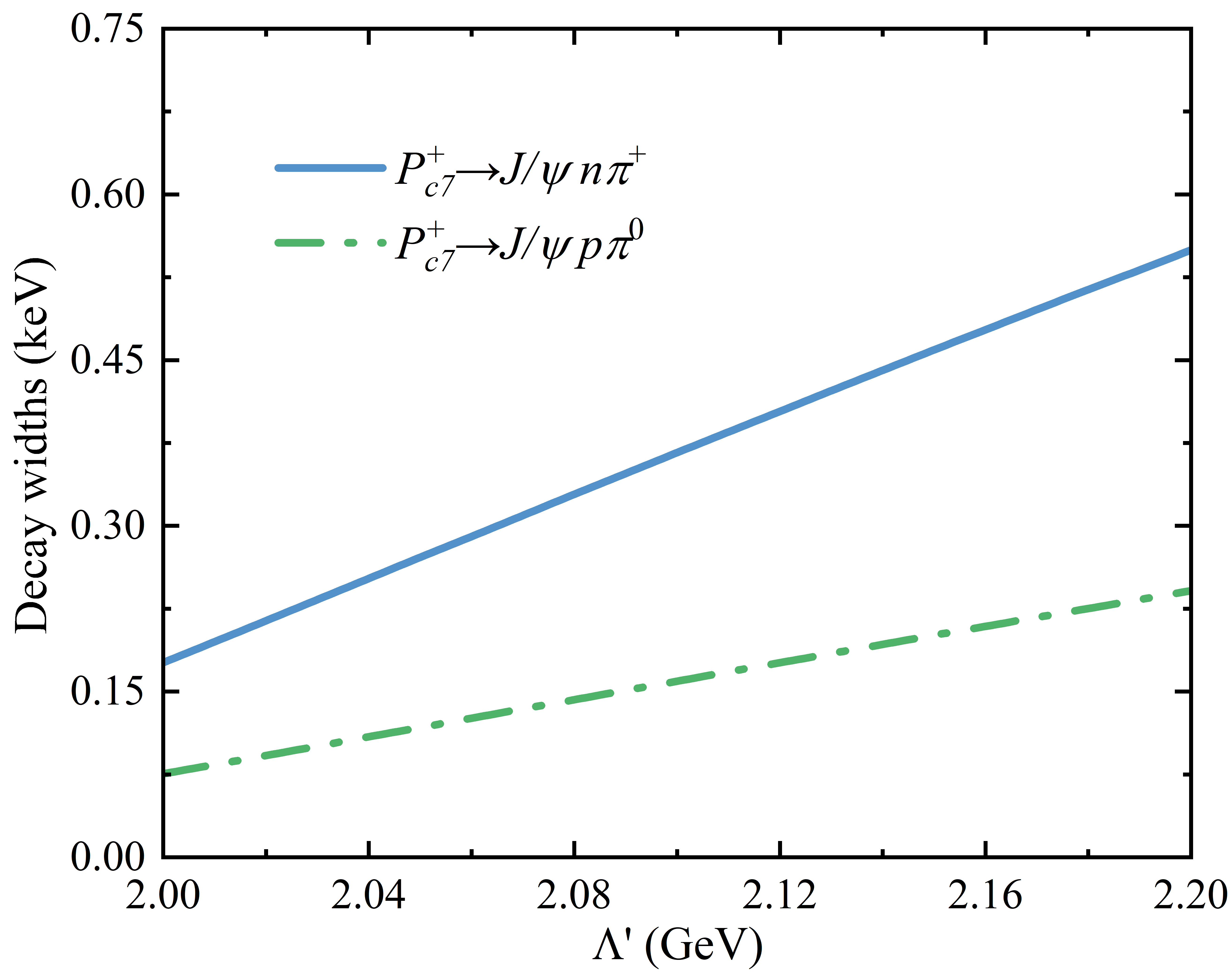}
}
\quad
\subfigure[\ $P_{c7}$-Scenario B]{
\includegraphics[width=7.2cm]{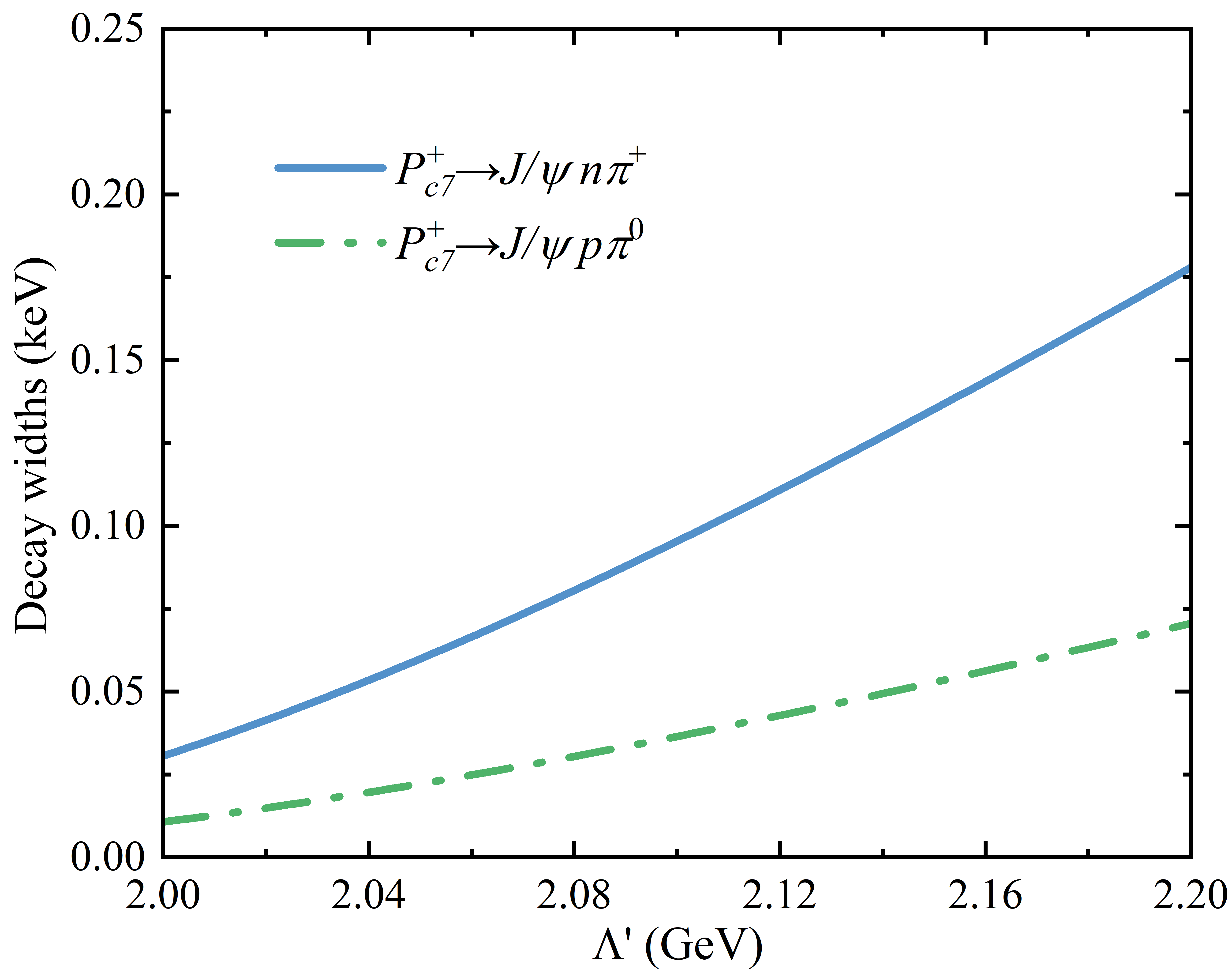}
}
\caption{ Partial decay widths of  $P_{c5|c6|c7}^+ \to J/\psi n \pi^+$ and $P_{c5|c6|c7}^+ \to J/\psi p \pi^0$ as a function of the cutoff $\Lambda'$ of the form factor (\ref{form factor}).   The results  are represented  by the blue solid lines and green dot-dashed lines, respectively, which correspond to the couplings determined for a cutoff  $\Lambda=$1.5 GeV.  }
\label{pc567}
\end{figure}

Apart from decaying into $J/\psi N \pi$,  $P_{c}(4440)$ and $P_{c}(4457)$ can decay into $\eta_{c} N \pi$ via 
$\bar{D}\Sigma_{c}$ {rescattering} to $\eta_{c}N$ if they have $J=1/2$.  It is natural to expect that the partial decay widths of $P_{c}(4440)$ and $P_{c}(4457)$  into $\eta_{c} N \pi$  are larger than those into $J/\psi N \pi$ because $P_{c}(4312)$ couples more to  $\eta_{c} N$  than to $J/\psi N$. 
In Fig.~\ref{Pc34-eta}, we plot the partial decay widths of $P_{c}(4440)$ and $P_{c}(4457)$  into $\eta_{c} N \pi$ as a function of the cutoff parameter $\Lambda'$. They are tens of keV, which are much larger than those  into $J/\psi N \pi$ in Fig.~\ref{Pc34}.   However, the widths in both Scenario A and  B are similar, which also cannot discriminate the spins of $P_{c}(4440)$ and $P_{c}(4457)$.  On the other hand, the branching ratios of  $P_{c}(4440)$ and $P_{c}(4457)$ decaying into  $\eta_{c} N \pi$, $J/\psi N \pi$, and $\bar{D}\Lambda_{c}\pi$ obtained in this work can be used to test the molecular nature of $P_{c}(4440)$ and $P_{c}(4457)$ if these decay channels are detected in future experiments.

We further predict the partial decay widths of $P_{c5}$,  $P_{c6}$, and $P_{c7}$  into $J/\psi N \pi$ in the  triangle-loop mechanism. In Fig.~\ref{pc567}, we plot their widths as a function of  the cutoff parameter $\Lambda'$. The partial decay widths of $P_{c5}$ and $P_{c6}$ in Scenario B are {  several keV}, while those in Scenario A are {  at the order of 1 keV}. The partial decay widths of $P_{c5}$ in Scenario A and B are more different than those of $P_{c6}$ due to the larger difference in the phase space of $P_{c5}$ into $J/\psi N \pi$ as shown in  Table~\ref{tab:coupling1} and Table~\ref{tab:coupling2}, similar to the case  of $P_{c3}$.  The partial widths of $P_{c7}$ decaying into $J/\psi N \pi$ are less than 1 keV, while the width in  Scenario A is larger than that in Scenario B, similar to the case of $P_{c4}$. 
The partial widths of  $P_{c7}$ decaying into  $J/\psi N \pi$ are much smaller than those of  $P_{c5}$ and  $P_{c6}$, which are suppressed by the higher spin of $P_{c7}$, the smaller  phase space of $P_{c7}$ decaying into $J/\psi N \pi$ in Scenario B, and the smaller coupling of $P_{c7}$ to $\bar{D}^{\ast}\Sigma_{c}^{\ast}$ in Scenario A.

\section{Summary and Discussion}
\label{sum}

Inspired by the discovery of $T_{cc}$ by the LHCb Collaboration in  the $DD\pi$ three-body final state, we  adopted the effective Lagrangian approach to systematically study seven $\bar{D}^{(\ast)}\Sigma_{c}^{(\ast)}$ hadronic molecules  decaying into three-body final states via two modes, tree level   and triangle loop. In the tree-level mode  the $\bar{D}^{(\ast)}\Sigma_{c}^{(\ast)}$  molecules decay via subsequent decays of  $\Sigma_{c}$ and $\Sigma_{c}^{\ast}$ into $\Lambda_c \pi$.  In the triangle mode the $\bar{D}^{(\ast)}\Sigma_{c}^{\ast}$ molecules decay into $J/\psi N \pi$ and $\eta_{c} N \pi $ via a two-step process, i.e., first $\bar{D}^{\ast}$ decays into $\bar{D}\pi$, and then $\bar{D}\Sigma_{c}^{(\ast)}$ rescatters  into $J/\psi N$ and $\eta_{c}N$.  The masses and widths of the $\bar{D}^{(\ast)}\Sigma_{c}^{(\ast)}$ molecules and relevant couplings  were determined in the coupled-channel contact-range EFT approach.

Our results show that the partial decay widths of $P_{c2}$, $P_{c}(4457)$, $P_{c5}$, $P_{c6}$ and $P_{c7}$  into $\bar{D}^{(\ast)}\Lambda_{c}\pi$, of the order of several MeV, are much larger than those of $P_{c}(4312)$ and $P_{c}(4440)$, and therefore are more accessible in future experiments.  The partial decay widths of  $P_{c}(4440)$ and $P_{c}(4457)$  into $J/\psi N \pi$ and $\eta_{c} N \pi$ are only several and tens of keV, respectively,  both of which are similar in scenarios A and B. We predicted the partial decay widths of $P_{c5}$, $P_{c6}$, and $P_{c7}$ into $J/\psi N \pi$, among which the width of $P_{c7} \to J/\psi N \pi$ is one order of magnitude smaller  than those of $P_{c5}$  and $P_{c6}$. Our results suggest that one should look for $P_{c7}$ in the $J/\psi N \pi$ and $\bar{D}^{\ast}\Lambda_{c}\pi$ invariant mass distributions, while the latter is preferable.  
These three-body  decay modes of the pentaquark states  are of great value to further observations of the pentaquark states. In addition,  those partial decay widths are helpful to test their molecular nature. 

It should be noted that although the predicted partial decay widths of $P_{c2}$, $P_{c5}$, $P_{c6}$, $P_{c7}$ are all dependent on the adopted value for the coupling $g$, our qualitative conclusions should be relatively robust, unless HQSS is broken much strongly than naively anticipated. As a result, the present study of three-body decay modes is expected to stimulate future experimental searches for the known and predicted $\bar{D}^{(*)}\Sigma_c^{(*)}$ molecules.

\section{Acknowledgments}
MZL thank Jun-Xu Lu and Ya-Wen Pan  for useful discussions. 
  This work is supported in part by the National Natural Science Foundation of China under Grants No.11975041,  No.11735003, and No.11961141004. Ming-Zhu Liu acknowledges support from the National Natural Science Foundation of
China under Grant No.12105007 and    China Postdoctoral
Science Foundation under Grants No. 2022M710317, and No. 2022T150036.

\appendix

\section{Invariant amplitudes for the tree-level and triangle-loop processes}
\label{appendix B}
The tree-level amplitudes of the $\bar{D}^{(\ast)}\Sigma_{c}^{(\ast)}$ hadronic molecules   decaying  into $\bar{D}^{(\ast)}\Lambda_{c}\pi$ read
\begin{eqnarray}
\mathcal{M}_{P_{c1}\to \bar{D}\Lambda_{c}\pi}&=& i\frac{g_{P_{c1}\bar{D}\Sigma_c}g_{\Sigma_{c}\Lambda_{c}\pi}}{f_{\pi}}~\bar{u}(p_{1}) {p\!\!\!/}_{2} \gamma_{5} \frac{1}{{k\!\!\!/}_{1}-m_{\Sigma_{c}}}u(k_{0}), \\ 
\mathcal{M}_{P_{c2}\to \bar{D}\Lambda_{c}\pi}&=& i\frac{g_{P_{c2}\bar{D}\Sigma_c^{\ast}}g_{\Sigma_{c}^{\ast}\Lambda_{c}\pi}}{f_{\pi}}~\bar{u}(p_{1}) {p}_{2 \mu} \frac{{k\!\!\!/}_{1}+m_{\Sigma_{c}^{\ast}}}{k_1^2-m_{\Sigma_{c}^{\ast}}^2+im_{\Sigma_{c}^{\ast}}\Gamma_{\Sigma_{c}^{\ast}}}P^{\mu\nu}(k_1)u_{\nu}(k_{0}), \\
\mathcal{M}_{P_{c3}\to \bar{D}^{\ast}\Lambda_{c}\pi}&=& i\frac{g_{P_{c3}\bar{D}^{\ast}\Sigma_c}g_{\Sigma_{c}\Lambda_{c}\pi}}{f_{\pi}}~\bar{u}(p_{1}) {p\!\!\!/}_{2} \gamma_{5} \frac{1}{{k\!\!\!/}_{1}-m_{\Sigma_{c}}}\widetilde{\gamma}^{\mu}\gamma_{5}\varepsilon_{\mu}(p_3)u(k_{0}), \\
\mathcal{M}_{P_{c4}\to \bar{D}^{\ast}\Lambda_{c}\pi}&=& i\frac{g_{P_{c4}\bar{D}^{\ast}\Sigma_c}g_{\Sigma_{c}\Lambda_{c}\pi}}{f_{\pi}}~\bar{u}(p_{1}) {p\!\!\!/}_{2} \gamma_{5} \frac{1}{{k\!\!\!/}_{1}-m_{\Sigma_{c}}}\varepsilon_{\mu}(p_3)u^{\mu}(k_{0}),  \\
\mathcal{M}_{P_{c5}\to \bar{D}^{\ast}\Lambda_{c}\pi}&=& i\frac{g_{P_{c5}\bar{D}^{\ast}\Sigma_c^{\ast}}g_{\Sigma_{c}^{\ast}\Lambda_{c}\pi}}{f_{\pi}}~\bar{u}(p_{1}) p_{2 \mu} \frac{{k\!\!\!/}_{1}+m_{\Sigma_{c}^{\ast}}}{k_1^2-m_{\Sigma_{c}^{\ast}}^2+im_{\Sigma_{c}^{\ast}}\Gamma_{\Sigma_{c}^{\ast}}}P^{\mu\nu}(k_1)\varepsilon_{\nu}(p_3)u(k_{0}),  \\
\mathcal{M}_{P_{c6}\to \bar{D}^{\ast}\Lambda_{c}\pi}&=& i\frac{g_{P_{c6}\bar{D}^{\ast}\Sigma_c^{\ast}}g_{\Sigma_{c}^{\ast}\Lambda_{c}\pi}}{f_{\pi}}~\bar{u}(p_{1}) p_{2\mu} \frac{{k\!\!\!/}_{1}+m_{\Sigma_{c}^{\ast}}}{k_1^2-m_{\Sigma_{c}^{\ast}}^2+im_{\Sigma_{c}^{\ast}}\Gamma_{\Sigma_{c}^{\ast}}}P^{\mu\nu}(k_1)\varepsilon_{\rho}(p_3)\gamma_5 \widetilde{\gamma}^{\rho}u_{\nu}(k_{0}),  \\
\mathcal{M}_{P_{c7}\to \bar{D}^{\ast}\Lambda_{c}\pi}&=& i\frac{g_{P_{c7}\bar{D}^{\ast}\Sigma_c^{\ast}}g_{\Sigma_{c}^{\ast}\Lambda_{c}\pi}}{f_{\pi}}~\bar{u}(p_{1})p_{2\mu} \frac{{k\!\!\!/}_{1}+m_{\Sigma_{c}^{\ast}}}{k_1^2-m_{\Sigma_{c}^{\ast}}^2+im_{\Sigma_{c}^{\ast}}\Gamma_{\Sigma_{c}^{\ast}}}P^{\mu\nu}(k_1)\varepsilon_{\rho}(p_3){u_{\nu}}^{\rho}(k_{0}), 
\end{eqnarray}
where  $k_{0}$ and $k_{1}$ denote the momenta of initial states and intermediate states, and the momenta of  $\Lambda_{c}$,  $\pi$, and $\bar{D}^{(\ast)}$ are represented by  $p_{1}$,  $p_{2}$, and $p_{3}$, respectively.  $\bar{u}(p_{1})$  and  $u(k_{0})$ denote the final and initial spinor wave functions, respectively. $P^{\mu\nu}(p)=g^{\mu\nu}-\frac{1}{3}\gamma^{\mu}\gamma^{\nu}-\frac{\gamma^{\nu}p^{\mu}-\gamma^{\mu}p^{\nu}}{3m}-\frac{2p^{\mu}p^{\nu}}{3m^{2}}$ denotes the propagator of a massive particle of spin $3/2$.

The amplitudes of  the $\bar{D}^{\ast}\Sigma_{c}$ molecules  decaying into $\eta_{c} N \pi$  read
 \begin{eqnarray}
 \nonumber
 i\mathcal{M}_{1/2}&=&g_{P_{c3}\bar{D}^{\ast} \Sigma_{c}} g_{D^{\ast} D \pi}\int \frac{d^{4}q}{(2\pi)^4}  \bar{u}(p_{1}) T_{\bar{D}\Sigma_{c}\to \eta_{c} N}(\sqrt{s})\frac{1}{{/\!\!\!k}_{1}-m_{\Sigma_{c}}}  \\  &&  \frac{1}{q^{2}-m_{\bar{D}}^{2}}p_{3\alpha}\frac{-g^{\alpha\beta}+\frac{k_{2}^{\alpha}k_{2}^{\beta}}{m_{\bar{D}^{\ast}}^2}}{k_{2}^2-m_{\bar{D}^{\ast}}^2}\widetilde{\gamma}_{\beta}\gamma_{5}u(k_0)F(q^2), \\
 \nonumber
 i\mathcal{M}_{3/2}&=&g_{P_{c4}\bar{D}^{\ast} \Sigma_{c}} g_{D^{\ast} D \pi} \int \frac{d^{4}q}{(2\pi)^4}  \bar{u}(p_{1})  T_{\bar{D}\Sigma_{c}\to \eta_{c} N}(\sqrt{s}) \frac{1}{{/\!\!\!k}_{1}-m_{\Sigma_{c}}} \\  && \frac{1}{q^{2}-m_{\bar{D}}^{2}}p_{3\alpha}\frac{-g^{\alpha\beta}+\frac{k_{2}^{\alpha}k_{2}^{\beta}}{m_{\bar{D}^{\ast}}^2}}{k_{2}^2-m_{\bar{D}^{\ast}}^2}u_{\beta}(k_0)F(q^2),
 \end{eqnarray}
and the amplitudes of  the $\bar{D}^{\ast}\Sigma_{c}^{(\ast)}$ molecules  decaying into $J/\psi N \pi$  read
 \begin{eqnarray}
\nonumber
 i\mathcal{M}_{1/2}&=&g_{P_{c3}\bar{D}^{\ast} \Sigma_{c}} g_{D^{\ast} D \pi}\int \frac{d^{4}q}{(2\pi)^4}  \bar{u}(p_{1}) \varepsilon_{\mu}(p_2)\gamma^{\mu}\gamma_{5}T_{\bar{D} \Sigma_{c} \to J/\psi N}(\sqrt{s})\frac{1}{{/\!\!\!k}_{1}-m_{\Sigma_{c}}} \\
 &&  \frac{1}{q^{2}-m_{\bar{D}}^{2}}p_{3\alpha}\frac{-g^{\alpha\beta}+\frac{k_{2}^{\alpha}k_{2}^{\beta}}{m_{\bar{D}^{\ast}}^2}}{k_{2}^2-m_{\bar{D}^{\ast}}^2}\widetilde{\gamma}_{\beta}\gamma_{5}u(k_0)F(q^2),  \\ 
 \nonumber
 i\mathcal{M}_{3/2}&=&g_{P_{c4}\bar{D}^{\ast} \Sigma_{c}}  g_{D^{\ast} D \pi} \int \frac{d^{4}q}{(2\pi)^4}  \bar{u}(p_{1})\varepsilon_{\mu}(p_2)\gamma^{\mu}\gamma_{5} T_{\bar{D} \Sigma_{c} \to J/\psi N}(\sqrt{s})  \frac{1}{{/\!\!\!k}_{1}-m_{\Sigma_{c}}} \\  && \frac{1}{q^{2}-m_{\bar{D}}^{2}}p_{3\alpha}\frac{-g^{\alpha\beta}+\frac{k_{2}^{\alpha}k_{2}^{\beta}}{m_{\bar{D}^{\ast}}^2}}{k_{2}^2-m_{\bar{D}^{\ast}}^2}u_{\beta}(k_0)F(q^2),\\
 \nonumber
  i\mathcal{M}_{1/2}&=&g_{P_{c5}\bar{D}^{\ast} \Sigma_{c}^{\ast}}  g_{D^{\ast} D \pi}\int \frac{d^{4}q}{(2\pi)^4}  \bar{u}(p_1)  T_{\bar{D} \Sigma_{c}^{\ast} \to J/\psi N} (\sqrt{s})\varepsilon_{\nu}(p_2)\\  && \frac{{k\!\!\!/}_{1}+m_{\Sigma_{c}^{\ast}}}{k_1^2-m_{\Sigma_{c}^{\ast}}^2+im_{\Sigma_{c}^{\ast}}\Gamma_{\Sigma_{c}^{\ast}}}{P^{\nu}}_{\beta}(k_1) \frac{1}{q^{2}-m_{\bar{D}}^{2}}p_{3\alpha}  \frac{-g^{\alpha\beta}+\frac{k_{2}^{\alpha}k_{2}^{\beta}}{m_{\bar{D}^{\ast}}^2}}{k_{2}^2-m_{\bar{D}^{\ast}}^2}u(k_0)F(q^2),  \\ 
 \nonumber
  i\mathcal{M}_{3/2}&=&g_{P_{c6}\bar{D}^{\ast} \Sigma_{c}^{\ast}}g_{D^{\ast} D \pi}\int \frac{d^{4}q}{(2\pi)^4}  \bar{u}(p_1) T_{\bar{D} \Sigma_{c}^{\ast} \to J/\psi N} (\sqrt{s}) \varepsilon_{\nu}(p_2) \\  &&  \frac{{k\!\!\!/}_{1}+m_{\Sigma_{c}^{\ast}}}{k_1^2-m_{\Sigma_{c}^{\ast}}^2+im_{\Sigma_{c}^{\ast}}\Gamma_{\Sigma_{c}^{\ast}}}P^{\nu\rho}(k_1)\frac{1}{q^{2}-m_{\bar{D}}^{2}}p_{3\alpha}  \frac{-g^{\alpha\beta}+\frac{k_{2}^{\alpha}k_{2}^{\beta}}{m_{\bar{D}^{\ast}}^2}}{k_{2}^2-m_{\bar{D}^{\ast}}^2}\gamma_{5}\widetilde{\gamma}_{\beta}u_{\rho}(k_0)F(q^2),  \\
  \nonumber
  i\mathcal{M}_{5/2}&=&g_{P_{c7}\bar{D}^{\ast} \Sigma_{c}^{\ast}}g_{D^{\ast} D \pi}\int \frac{d^{4}q}{(2\pi)^4}  \bar{u}(p_1) T_{\bar{D} \Sigma_{c}^{\ast} \to J/\psi N} (\sqrt{s}) \varepsilon_{\nu}(p_2) \\  &&  \frac{{k\!\!\!/}_{1}+m_{\Sigma_{c}^{\ast}}}{k_1^2-m_{\Sigma_{c}^{\ast}}^2+im_{\Sigma_{c}^{\ast}}\Gamma_{\Sigma_{c}^{\ast}}}P^{\nu\rho}(k_1)\frac{1}{q^{2}-m_{\bar{D}}^{2}}p_{3\alpha}  \frac{-g^{\alpha\beta}+\frac{k_{2}^{\alpha}k_{2}^{\beta}}{m_{\bar{D}^{\ast}}^2}}{k_{2}^2-m_{\bar{D}^{\ast}}^2}u_{\beta\rho}(k_0)F(q^2), 
 \label{pi}
 \end{eqnarray}
 where $q$, $k_{0}$, $k_{1}$,  $k_{2}$,  $p_{1}$, $p_{2}$ and $p_{3}$ represent the momenta of $\bar{D}$, $P_{c}$, $\Sigma_{c}^{(\ast)}$, $\bar{D}^{\ast}$, $N$, $J/\psi(\eta_c)$ and  $\pi$, respectively, and $T_{\bar{D}\Sigma_{c}^{(\ast)}-J/\psi(\eta_c) N}$ represent the  potentials  of inelastic scattering. To avoid the divergence induced by the loop function, we have introduced  a form factor of the form 
 \begin{equation}
     F(q^2)=\left(\frac{\Lambda'^{2}-m_{\bar{D}}^2}{\Lambda'^{2}-q^2}\right)^2.
     \label{form factor}
 \end{equation}

\bibliography{XiccSigmac}

\end{document}